\magnification=1200
%\magnification=\magstep1
\input amstex
\documentstyle{amsppt}
%\UseAMSsymbols
\voffset=-3pc
\loadbold
\loadmsbm
\loadeufm
%\UseAMSsymbols
%\baselineskip=12pt
%\parskip=6pt
\def\var{\varepsilon}
\def\bC{\Bbb C}

\def\bR{\Bbb R}

\def\bZ{\Bbb Z}

\def\cA{\Cal A}

\def\fa{\frak a}
\def\fg{\frak g}
\def\fk{\frak k}
\def\fh{\frak h}
\def\fu{\frak u}
\def\fp{\frak p}

\def\Im{\text{Im}\,}
\def\mt{\mapsto}
\def\ra{\rightarrow}

\def\ad{\text{\rm ad}}
\def\sh{\text{\rm sh}}

%\NoBlackBoxes
\topmatter
\title{Quantization of compact Riemannian symmetric spaces}
\endtitle
\author  R\'obert Sz\H{o}ke
%\footnote {This research was partially supported by 
%  OTKA grant}
% N81203.} 
\endauthor
\rightheadtext{ R\'obert Sz\H{o}ke}
\leftheadtext{Quantization of compact Riemannian symmetric spaces}
\address
%\hskip-.20truein
 Department of Analysis, Institute of Mathematics, ELTE E\"otv\"os Lor\'and
 University, 
P\'azm\'any P\'eter s\'et\'any 1/c, Budapest 1117, Hungary,
\newline ORCID ID:0000-0002-8723-1068
\email rszoke\@cs.elte.hu
\endemail
\endaddress
%\acknowledgement We are
%\endacknowledgement
\keywords
adapted complex structures, geometric quantization, Hilbert fields
\endkeywords 
\subjclassyear{2000}
\subjclass 53D50, 53C35, 32L10, 70G45,65
\endsubjclass
\abstract
The phase
space of  a compact, irreducible,  simply connected,   Riemannian 
symmetric space admits a natural  family of  K\"ahler
polarizations parametrized by the upper half plane $S$. Using this
family,
geometric quantization, including the 
half-form correction, produces 
 the  field $H^{corr}\ra S$
of quantum Hilbert spaces.  
 We show
  that  projective flatness of $H^{corr}$  
implies, that the symmetric space must be isometric to a compact Lie
group equipped with a biinvariant metric.
In the latter case the flatness of  $H^{corr}$  was 
previously established. 

\endabstract
\endtopmatter
\document
\subhead 0. Introduction
\endsubhead
Suppose the 
configuration space of a classical mechanical system is
an $m$--dimensional compact Riemannian manifold $M$ and 
the metric corresponds to 
twice the kinetic energy.
The aim of geometric quantization is to construct a Hilbert space
 (the quantum Hilbert space) associated to
this system, in a natural way.

According to the prescriptions of Kostant and Souriau 
[Ko1,So,Wo],
the first step in this process,
   is to pass to phase space $(N,\omega)$, which for the moment we take
   $TM\approx T^*M$, 
a symplectic manifold with an exact symplectic form, and then to choose
a Hermitian line bundle with connection $E\to N$,
the so called prequantum
line bundle,   whose curvature is $-i\omega$.
This bundle is unique when $M$ is  simply connected.
%$N$ can be taken for example 
% $TM\approx T^*M$,
%with its standard exact symplectic form 
%$\omega$, that equals to $\sum dq_j\wedge dp_j$
%in the usual local coordinates.
  
%If $M$ is simply connected, the bundle is unique up to a connection 
%preserving Hermitian isomorphism.
%In any case, one such line bundle is obtained from a real 1--form $a$ on $N$ 
%such that $da=-\omega$, by letting $E=N\times\bC\to N$ to be the trivial 
%line bundle with $h^E(x,\gamma)=|\gamma|^2$ the trivial metric on it. 
%If sections are
%identified with functions $\psi\colon N\to\bC$,
%the connection $\nabla^E$ is defined by
%$$
%\nabla_\zeta^E\psi=\zeta\psi+ia(\zeta)\psi,\qquad\zeta\in\Vect \,N.
%$$

The next step is a  choice of a   K\"ahler  structure on $N$
with K\"ahler form $\omega$. This  induces
on $E$ the structure of a 
holomorphic line bundle  and gives rise to the quantum Hilbert space $H$,
consisting of holomorphic sections of $E$ that are $L^2$ with
respect to the volume form $\omega^m/m!\,$.
Often one  includes in this 
construction the so called half-form correction. Suppose
 $\kappa$ is a square root of the canonical
 bundle $K_N$. Then the corrected quantum Hilbert space $H^{corr}$ consists of 
the $L^2$ holomorphic sections of $E\otimes\kappa$.

The quantum Hilbert space obtained this way depends on the
choices made in this process %Even if we restrict this choice, by
%imposing a reasonable condition of the nature that the K\"ahler structure
%should respect the symmetries present, the structure will still not be
%unique,
and the question arises whether there is a canonical way to
identify the quantum Hilbert spaces corresponding to the different choices.
This question
is a fundamental issue in geometric quantization, 
that has been studied from different perspectives in several papers, see e.g. 
[ADW, Bl1-2, Ch, F, FMN1-2, FU, Hal1-2, Hi, Ko2, KW, OW, R, S, Vi].

When $M$ is a real-analytic Riemannian manifold, there is  a natural 
K\"ahler polarization defined at least in  some
neighborhood $X\subset TM$ of the zero section  of $N$ ([GS, HK, Sz1]),
the so called {\it adapted complex structure},  
in which $\omega$ becomes a K\"ahler form. 
(See Sect. 1.1 for more details on adapted complex structures.)
In good cases $X=N$. One gets examples of this sort  when
 $M$ is a compact Riemannian symmetric space or more generally a
 compact normal
Riemannian homogeneous space [Sz1, Sz3],   
but there are nonhomogeneous examples as well, see [A].

In fact the adapted complex structure is just one member
 in a natural family of K\"ahler structures on $N$ [L-Sz2],
 that is parametrized by the upper half plane $S$.
This is the family of K\"ahler structures that respects the symmetries of
$N$. Here Im$s$, $s\in S$ plays the role of
Planck's constant (cf. Sect. 1.1).

Suppose  
for the compact Riemannian manifold $M$, the adapted complex
structures $J(s)$, $s\in S$  exist on $N$. 
Geometric quantization then 
produces a family of quantum Hilbert spaces. Our main concern is how (and
 when) can one define a natural (projective) isomorphism among these Hilbert
spaces.

Whenever one has a family of K\"ahler structures on the phase space $N$
pa\-ra\-met\-ri\-zed by some smooth manifold $S$,  
  [ADW] and [Hi] suggest  to view the corresponding 
    collection $\{H_s : s\in S\}$ (or 
$H^{corr}_s$) of Hilbert spaces 
as the fibers of a   Hilbert bundle $H\ra S$ (resp. $H^{corr}\ra S$)
endowed with some Hermitian 
connection, the quantum connection.
If this were true,
parallel transport along a curve in
$S$ would yield a unitary map between different fibers.
If the quantum  connection were  
projectively flat, and $S$ simply connected,  parallel translation 
 would even yield a path-independent canonical projective unitary map between
 $H_{s_1}$ and $H_{s_2}$ for arbitrary parameters $s_1, s_2\in S$.
When $S$ is a complex
manifold and the bundle $H\ra S$ is holomorphic,
there is a canonical choice of  a Hermitian connection,
the Chern connection.

% For this purpose one needs to define a smooth (complex) manifold structure
 %on $H$ ($H^{corr}$) and should show that it is locally trivial.

Now  in the uncorrected version    
  one may try to implement the idea above as follows.   
 For each $s\in S$, $H_s$ is
 a closed Hilbert subspace of $K$, the $L^2$ sections of the
 smooth prequantum line bundle $E\ra N$. 
 Thus we could view   the set  $H$, the set theoretical disjoint union
 of all the $H_s$,   as a subset of the trivial Hilbert
 bundle $pr:\Cal K=S\times K\ra S$, such that $H_s$ is a closed Hilbert
 subspace
  of the fiber $pr^{-1}(s)$ for each $s\in S$.
This way $H$ certainly inherits a topology from $\Cal K$. 
It is much less clear
 whether $H$ inherits (or under what conditions)
 a  complex (or even a smooth)  manifold structure. It is even less clear
 why (and when) $H$ is a subbundle (of some kind) of $\Cal K$.

The corrected version  is more complicated. Now in addition to the problems
we faced in the uncorrected case, an extra complication arises.
The corrected quantum
Hilbert spaces $H^{corr}$  are no longer subspaces of a fixed Hilbert space,
but rather of a varying family $K_s$ of Hilbert spaces, where $K_s$ denotes
the $L^2$ sections of the corrected prequantum  line bundle
$E\otimes\kappa_s$.

It is  not clear if the $K_s$ family itself
forms a holomorphic (smooth)
Hilbert bundle $\Cal K\ra S$ or not (whether $H^{corr}\ra S$ is a
smooth (holomorphic) subbundle of $\Cal K$).

In fact it turns out that there are at least two 
equally natural but inequivalent ways to make $K_s$ a smooth Hilbert
bundle. That means  on this family
there are two natural and different smooth Hilbert bundle structures
(see [Sz4]).

To avoid these difficulties it is better not  to put
any smooth structure on an object  like $H^{corr}\ra S$ initially,
rather  try to understand its structure through its sections. 
This was the motivation for us to 
  introduce in [LSz3] the notion of a smooth or analytic
field of Hilbert spaces,
generalizing Hilbert bundles that [ADW] worked with.

A field of Hilbert spaces is  simply
a map $p:H\rightarrow S$ of sets with each
fiber $H_s=p^{-1}(s)$ endowed with the structure of a Hilbert space.
Since we do not put any topology on $H$, a section of $p:H\rightarrow S$
simply means any map $\varphi:S\ra H$ with $\varphi(s)\in H_s$.

When $S$ is a manifold, one says that a smooth (resp. analytic)
 structure on $H$ is
 specified (with which $H$ becomes a smooth [analytic]
 field of Hilbert spaces) if
 a vector space $\Gamma^\infty$ (resp. $\Gamma^\omega$)
 of sections of $H$ is chosen, together with a connection
 like operation on it in such a way that they satisfy a natural set of axioms
 imitating  as if $\Gamma^\infty$ (resp. $\Gamma^\omega$)
 was the vector space of  smooth (real-analytic) sections
 of a smooth (real-analytic) Hilbert bundle equipped with a Hermitian connection
 (see [LSz3] and Sect. 1.2 for the precise definition and more details on
 fields of Hilbert spaces).

Although these objects are quite a bit more general than ordinary
Hilbert bundles, still with the help of the
connection-like operation built into their definition,
it still makes sense to talk about its curvature.
Similarly to the classical situation if the Hilbert field is analytic and the
curvature turns out to be $0$, or at least ``central" (=projectively flat),
then path
independent parallel transport  allows for canonical identification of
the (projectivized) quantum Hilbert spaces [LSz3, Theorem 2.3.2,
Theorem 2.4.2].

When
$M$ is a  
compact,  normal 
Riemannian homogeneous space,
%with the help of the  complex structure of $Y$
one can 
naturally 
 endow the field
$H^{corr}\ra S$ with an  analytic structure
  ([L-Sz3, Theorem 11.1.1]).
Our main result is:

\proclaim{Theorem 0.1}
Let $(M,g)$ be a  compact, irreducible, simply connected,   Riemannian
 symmetric space. Assume the corrected field of quantum Hilbert spaces 
$H^{corr}\ra S$ is 
projectively flat. Then 
$M$  is  isometric to a group manifold
(i.e. a  compact, connected, simple, simply connected
Lie group  equipped with a biinvariant metric).
\endproclaim

We prove Theorem 0.1 in Sect. 6.
Group manifolds were treated in [Lsz3, Theorem 11.3.1],
where 
it was shown 
 that whenever $M$ is isometric to  a compact,
simply connected Lie group with a biinvariant metric, 
 the field $H^{corr}\ra S$ is 
flat.
Together with Theorem 0.1 we get Corollary 0.2.

\proclaim{Corollary 0.2} Let $(M,g)$ be a  compact, irreducible,
simply connected,   Riemannian
 symmetric space. Then the corrected field of quantum Hilbert spaces 
$H^{corr}\ra S$ is 
projectively flat if and only if $M$ is isometric to 
a group manifold.
%a compact, connected, simple, simply connected
%Lie group $U$ equipped with a biinvariant metric.
 In the latter case the field $H^{corr}\ra S$ is flat.
 \endproclaim

 Corollary 0.2 shows that quantization is unique
for group manifolds and for other symmetric spaces quantization
does depend on the choice of the K\"ahler polarization.

Flatness implies that $H^{corr}\ra S$ is a genuine Hilbert bundle (trivial
 in this case). It is not known whether this is also true when the Hilbert
 field is not  projectively flat.

Theorem 0.1 generalizes  [LSz3, Theorem 12.1.1], that dealt with spheres, and
[Lsz4, Theorem 1.1] that treated  rank$-1$ symmetric spaces.

The main scheme of the proof of the theorem is based on the rank$-1$
case [L-Sz4],
but the situation  here is much more complicated.

 Writing $M$ in the usual $M=U/K$ form (see Sect. 1.3 for the notation),
  each irreducible $K-$spherical representation of $U$ gives rise to  a 
certain integral   on the positive Weyl chamber (2.2.6). The integrand
involves the corresponding $K-$spherical function and it also depends
on a real parameter $\tau$, that takes  arbitrary positive values.
Projective flatness of the Hilbert
field  is expressed as a simple relation among
  these integrals (Theorem 2.1.1 (b)).
Since the  explicit value of these integrals is not known,
one needs  other ways to test projective flatness.

The idea in the rank$-1$ case ([LSz4]) was to tend with the parameter to zero
  resp.  to infinity,  calculate the asymptotic behavior of our integrals
  and compare the information obtained this way with the
  relation that holds among the integrals corresponding to different spherical
  representations.
 In the rank$-1$ situation $K-$spherical
  functions  are quite explicit, they reduce
 to hypergeometric polynomials,  greatly simplifying
 the situation.

  In the higher rank case, the basic idea is the same, but the situation is
  more involved.
  We still want 
  to calculate the asymptotic behavior of those (now multivariable)
  integrals as the
  parameter tends to zero, resp. to infinity.
 Spherical functions now correspond to multivariable
  Jacobi polynomials associated to the restricted root system ([H], [HO1-2],
  [HS]) and they are much more complicated functions to calculate with.

The key observation here is that despite  this, their main contribution
  to the asymptotic behavior of our integrals (when $\tau\ra\infty$) is simple.
  The Jacobi polynomials are actually exponential polynomials, where each
  term corresponds to a weight of the given $K-$spherical representation.
   The main contribution comes only from one term, that corresponds to the
   highest weight. We even know the coefficient of this term, it is
   Harish-Chandra's $\bold c-$function.
This is the content of Proposition 3.1 and
Theorem 5.2. As a consequence, projective
flatness implies that a certain numerical quantity $Q(\delta)$ (see (5.7))
associated to every irreducible $K-$spherical
representation $\delta$,   that involves only the usual $\Gamma$ function,
the restricted
root system, the multiplicities  and
 the highest weight of $\delta$, in fact is independent of the representation
 (Theorem 5.5).

Finally the question, for which spaces  will this be true,
can be translated to a problem about abstract root systems with
multiplicities.
  This problem is treated in Theorem 6.2.2, after which the proof of
  Theorem 0.1
  easily follows.

The organization of our paper is the following.
  After an introductory section, where we shortly summarize the necessary 
 background, in Sect. 2
 we  discuss the curvature of the field $H^{corr}\ra S$ when the manifold $M$
 is a compact, irreducible, simply connected Riemannian symmetric space.
In Sect. 3 we discuss spherical functions briefly. 
 The asymptotics of our integrals as $\tau\ra 0$ (resp. $\tau\ra \infty$)
 is calculated in Sect. 4 (resp. Sect. 5). Sect. 6 is devoted to root systems
 with multiplicities and at the end we prove our main theorem.
%\newpage

\head 1. Basic notations 
\endhead

\subhead 1.1. Adapted complex structures
\endsubhead
%\medskip
Here we review  some important facts
on adapted complex structures mainly from [LSz2].
Following Souriau's philosophy ([So]), we 
define
the phase space $N$ of a compact Riemannian manifold not as  $TM\approx T^*M$,
but rather
 as the manifold of parametrized geodesics $x:\bR\ra M$. Any $t_0\in\bR$ 
induces a diffeomorphism $N\ni x\mt\dot x(t_0)\in TM$, and  the pull back of 
the  canonical 
symplectic form of $TM\approx T^*M$ is independent of $t_0$; we denote it by 
$\omega$. We  identify $M$ with the submanifold of zero speed geodesics in $N$.
Affine reparametrizations $t\mt a+bt$, $a, b\in\bR$, act on $N$ and define
a right action of the Lie semigroup $\cA$ of affine reparametrizations.  

 Given a complex manifold structure on $\cA$, 
a complex structure on $N$ 
is called {\it adapted} if for every $x\in N$ the orbit map 
$\cA\ni\sigma\mt x\sigma
\in N$
is holomorphic. 
An adapted complex structure on $N$ can exist only if
the initial complex structure on $\cA$ is left invariant. The left invariant 
complex structures on $\cA$ are parametrized by the points of 
$\bC\setminus\bR$. (The points of $\bR\subset\bC$ correspond to left invariant
real polarizations on $\cA$.)
For each $s\in \bC\setminus\bR$ and corresponding left
 invariant complex structure $I(s)$ on $\cA$, if an $I(s)$ adapted complex 
structure $J(s)$ exists on $N$, 
then this structure is unique and if $J(i)$ exists, then $J(s)$ also exists
 for all $s$ in $s\in \bC\setminus\bR$. This is the case for example when
 $(M,g)$ is a compact symmetric space or more generally, a compact,  normal 
Riemannian homogeneous space ([Sz1, Sz3]).
The original definition
 of adapted complex structures in [L-Sz1, Sz1] corresponds to the parameter 
$s=i$.

In fact the  K\"ahler manifolds $(N, J(s))$
are all biholomorphic to a fixed one,
$(N,J(i))$, but the biholomorphism maps the K\"ahler form $\omega$
to $\omega/\Im s$; this is the content of [LSz3, (10.3.4)].
Thus $\Im s$ plays the role of Planck's constant.

Suppose we are in the situation when $J(s)$ exists.
Denote by $\partial_s$, $\bar\partial_s$ the complex exterior derivations for
this structure, and $L(x)$ the square of the speed of the geodesic $x\in N$.
Then $i\omega=(\text{Im}s)\bar\partial_s\partial_sL$. In particular
$L$ is a potential of a positive (resp. negative) K\"ahler structure
with K\"ahler form $\omega$, when $s$ is in
the upper (resp. lower) half plane.
 When $s$ is a point in the lower half plane, the only holomorphic $L^2$ section
 of the quantum line bundle $E\ra (N,J(s))$ is the identically zero section.
 For this reason we are only interested   in the $J(s)$ structures when
 $s$ is an element of the upper half plane, denoted by $S$.

It is important that the
family of adapted complex structures $J(s)$, $s\in S$ on $N$ can all 
be put together to form a ``twistor space'' like 
holomorphic fibration $\pi:Y\ra S$, where 
the fibers
 $Y_s=\pi^{-1}s$ are biholomorphic to  $(N,J(s))$.
As a differentiable manifold, $Y=S\times N$, and the projection
$pr:Y\ra N$ realizes the biholomorphisms $Y_s\ra(N,J(s))$
([L-Sz2, Theorem 5]).
The pull back $\tilde\omega$ of $\omega$ by $pr$ 
 satisfies
$$
 i\tilde\omega:=\bar\partial\partial(L\text{Im s})\quad
  \text{on }Y.\tag1.1.1
$$

Assuming now that $M$ is simply connected, with the help of the fibration
$\pi:Y\ra S$
one can perform geometric quantization
simultaneously  ([LSz3]). 
%The object we get is  a field of Hilbert spaces.
%To quantize $(N,J(s))$ simultaneously,
Because of (1.1.1), $\tilde\omega$ is an exact $(1,1)$ form.
Therefore the unique Hermitian 
line bundle with connection $(E,h^E)\ra Y$ whose curvature is $-i\tilde\omega$
becomes a holomorphic line bundle.
The restriction of $E$ to $Y_s$ yields the prequantum line bundle corresponding
to  $(N,J(s),\omega)$. The restriction  of the form $\nu=\tilde\omega^m/m!$
to a fiber $Y_s$ is a volume form.
 The spaces of holomorphic
$L^2$-sections of $E|Y_s$ form the  Hilbert field 
$H\ra S$.

Since $M$ is simply connected, there is a unique Hermitian 
holomorphic line bundle $\kappa$ on $Y$ [LSz3, Sect. 10.4], so that 
$\kappa\otimes\kappa\approx K_\pi$ (the relative canonical bundle of $Y$ with 
 $K_\pi|Y_s$ being the canonical bundle of $Y_s$).
 Let $(E^{corr},h^{E^{corr}})=(E\otimes\kappa,h^E\otimes h^\kappa)$.
The spaces of holomorphic
$L^2$-sections of $E^{corr}|Y_s$ form the corrected Hilbert field 
$p:H^{corr}\ra S$, which at the moment is just a map of sets where all the fibers
have a Hilbert space structure. The complex structure
 of $Y$ and the holomorphic fibration $\pi:Y\ra S$ plays a crucial role
 in the construction  of the extra structures (smooth and analytic) we need
 on the field $H^{corr}\ra S$.
  This is discussed in the next section.

\subhead 1.2. Fields of Hilbert spaces
\endsubhead
%\medskip
 Here we review some notions and results from [LSz3] concerning fields of
 Hilbert spaces.
A field of Hilbert spaces is simply  a map $p:H\rightarrow S$ of sets with each
fiber $H_s=p^{-1}(s)$ endowed with the structure of a Hilbert space.
The inner products on the fibers, taken together, define a function
$$
h\colon H\oplus H\to\bC,\qquad\text{ where }\qquad
H\oplus H=\coprod_{s\in S}\ H_s\oplus H_s.
$$
If $v\in H$, we also write $h(v)$ for $h(v,v)$.
Hilbert fields naturally arise as direct images of 
holomorphic vector bundles.
Let $\pi:Y\rightarrow S$ be a surjective holomorphic submersion of
 finite dimensional complex manifolds,
 where we do not assume that $\pi$ is proper. 
Let $\nu$  be a smooth form on $Y$ that restricts to a volume form
on each fiber $Y_s=\pi^{-1}s$ and let $(E,h^H)\rightarrow Y$ be a Hermitian  
holomorphic vector bundle of finite rank. 
Denote by $H_s$  the Hilbert space of $L^2$ holomorphic sections of $E|Y_s$. 
 Then $H_s$ forms a Hilbert field $H\rightarrow S$.
An example of this sort is $H^{corr}\ra S$ from Sect. 1.1.

Direct images tend to have looser structures then bundles.
As it was hinted at in the introduction,
instead of  specifying some smooth structure on the total space $H$,
we try to understand their structure
through their sections.
\definition{Definition 1.2.1} Let $S$ be a smooth manifold.
A smooth structure on a field $H\to S$ of Hilbert spaces is given by
specifying a set $\Gamma^\infty$ of sections of $H$, closed under addition
and under multiplication by elements of $C^\infty(S)$, together with
linear operators $\nabla_\xi\colon\Gamma^\infty\to\Gamma^\infty$ for each
vector field $\xi$ on  $S$, such that for 
 $f\in C^\infty(S)$, $\varphi,\psi\in\Gamma^\infty$ and vector fields $\xi,\eta$
$$
\gather
\nabla_{\xi+\eta}=\nabla_\xi+\nabla_\eta,\ \nabla_{f\xi}=
f\nabla_\xi,\ \nabla_\xi (f\varphi)=(\xi f)\varphi+f\nabla_\xi\varphi;\\
h(\varphi,\psi)\in C^\infty(S)\text{ and }\xi h(\varphi,\psi)=
h(\nabla_\xi\varphi,\psi)+h(\varphi,\nabla_{\overline\xi}\psi);\tag1.2.1\\
\{\varphi(s)\colon\varphi\in\Gamma^\infty\}\subset H_s
\text{ is dense, for all }s\in S.\hskip1true in\tag1.2.2
\endgather
$$
\enddefinition
The collection $\nabla$ of the operators $\nabla_\xi$ is called a
connection on $H$. A field of Hilbert spaces together with a smooth structure
is called a smooth Hilbert field.
The  curvature of $H\ra S$ is defined by 
$$
R(\xi,\eta)=\nabla_\xi\nabla_\eta-\nabla_\eta\nabla_\xi-\nabla_{[\xi,\eta]}:
\Gamma^\infty\rightarrow\Gamma^\infty
$$ 
The field $H$ is called flat if $R=0$ and projectively flat if 
$R(\xi,\eta)$ acts by  multiplication by a 
function $r(\xi,\eta):S\rightarrow\bC$. In the latter case,
similarly to vector bundles, 
$r$ is in fact a smooth closed 2-form on $S$, and  a simple
twisting will reduce projectively flat smooth Hilbert fields to flat ones.

%Let $V$ be a Hilbert space. A
% trivial Hilbert bundle $H=S\times V\ra S$ with its canonical Hermitian
% connection and $\Gamma^\infty$ as its smooth sections yields a  trivial smooth
% Hilbert field.
As  [LSz3, Example 2.3.4] shows, flatness of a smooth Hilbert
 field does not guarantee  its local triviality.  Here a Hilbert field
with a smooth structure  $(H,\Gamma^\infty,\nabla)$
  is called trivial if there exist a fixed Hilbert space $V$, a fiber
  preserving and fiberwise unitary bijection $T:H\ra S\times V$, such that
for any $\varphi\in\Gamma^\infty$ and vector field $\xi$ along $S$, 
   $T\varphi$ will be a $C^\infty$ section of $S\times V\ra S$ and
  $T(\nabla_\xi\varphi)=\xi T\varphi$.
 %One needs a stronger notion that we called analytic Hilbert field.
 %$\Vect^\omega S\subset\Vect\,S$ for the
%Lie algebra of analytic vector fields.
\definition{Definition 1.2.2}Let $H\to S$ be a
smooth Hilbert field over a real-analytic manifold $S$.

(i)
A section $\varphi\in\Gamma^\infty$ is analytic if for any compact
$C\subset S$ and any finite set $\Xi$ of vector fields,
analytic in a neighborhood of $C$, there is an $\var>0$ such that
$$
\sup{\var^n\over n!}\ h(\nabla_{\xi_n}\ldots
\nabla_{\xi_1}\varphi)(s)^{1/2}<\infty,
$$
where the $\sup$ is taken over $n=0,1,\ldots,\xi_j\in\Xi$, and $s\in C$.
The set of analytic sections is denoted by $\Gamma^\omega\subset\Gamma^\infty$.

(ii)\ $H\to S$ is an {\it analytic Hilbert field} if
$\{\varphi(s)\colon\varphi\in\Gamma^\omega\}\subset H_s$ is
dense for all $s\in S$.
\enddefinition
An analytic and flat Hilbert field $H\ra S$ will be locally trivial
([LSz3, Theorem 2.3.2]) and so parallel transport can be introduced that
identifies the fibers locally in a canonical way.

For a surjective holomorphic submersion $\pi:Y\rightarrow S$
and  a Hermitian  
holomorphic vector bundle  $(E,h^H)\rightarrow Y$, under appropriate
conditions on $Y$ and $E$, the direct image
Hilbert field $H\ra S$ comes naturally 
 endowed with a smooth
 structure ([L-Sz3, Sect. 6-7]).
 In the problem of geometric quantization by
adapted complex structures, these conditions are known to be satisfied 
 when 
 $M$ is a 
compact, simply connected, normal 
Riemannian homogeneous space. In this case
$H^{corr}\ra S$
turns out to be analytic ([L-Sz3, Theorem 11.1.1]).

In the rest of this section we sketch the
basic idea of the  construction of the (quantum) connection on $H^{corr}\ra S$.
Recall from Sect.
1.1 the  holomorphic submersion $\pi:Y=S\times N\ra S$
and Hermitian holomorphic vector bundle $(E^{corr},h^{E^{corr}})
\rightarrow Y$.
Sections $\varphi$ of $H^{corr}\ra S$ are in one to one correspondence
 with sections $\Phi$ of the bundle
 $E^{corr}$ that are holomorphic and $L^2$ on each
 $Y_s$, the correspondence is $\Phi(y)=\varphi(\pi y)(y)$, for $y\in Y$.
 Write $\Phi=\hat\varphi$ and $\varphi=\check\Phi$ to indicate this
 correspondence.
 
 A lift of a smooth, vector field $\xi$ defined on $S$, is a vector field
 $\hat\xi$ on $Y$ such that $\pi_*\hat\xi(y)=\xi(\pi(y))$, for $y\in Y$. Lifts
 are not unique, but we can at least require that
if $\xi$ is of type $(1,0)$ or $(0,1)$, the lift should be of the same type.
If $\hat\xi^1$ and $\hat\xi^2$ are two lifts of the same vector field,
then $\hat\xi^1-\hat\xi^2$ will be vertical, i.e. tangential to the fibers
$Y_s$.

Denote by $\widetilde\nabla$ the Chern connection of $(E^{corr},h^{E^{corr}})$.
This implies in
particular that whenever $Z$ is a smooth, vertical vector field on $Y$ of type
$(0,1)$ and
$\Phi$ a smooth section of $E^{corr}$, whose restriction to each $Y_s$ is
holomorphic, we have
$$
\widetilde\nabla_Z\Phi=0.\tag1.2.3
$$
If   $\hat\xi$ is a lift
of a smooth vector field $\xi$ on $S$ of type
$(0,1)$ and  $Z$  as before, we have 
$$
\widetilde\nabla_Z(\widetilde\nabla_{\hat\xi}\Phi)=
\widetilde\nabla_{\hat\xi}(\widetilde\nabla_Z\Phi)+\widetilde\nabla_{[Z,\hat\xi]}
\Phi-i\tilde\omega(Z,\hat\xi)\Phi=0.\tag1.2.4
$$
This holds because each term on the right hand side of (1.2.4) is zero:
the first because of (1.2.3), the second because (as one easily computes)
$[Z,\hat\xi]$ will be also vertical and  of type $(0,1)$
and again we can use (1.2.3) with $Z$ replaced by $[Z,\hat\xi]$, finally
the last term vanishes because the form $\tilde\omega$ is of type $(1,1)$
 and $Z$ and $\hat\xi$ are both of type $(0,1)$.

Now (1.2.3) and (1.2.4) together imply the following.
 Let $\xi$ be a smooth vector field of type $(0,1)$ on $S$, $\hat\xi$
 an arbitrary lift (also of type $(0,1)$) to $Y$ and
 $\Phi$ a smooth section of $E^{corr}$ whose restriction to each $Y_s$ is
holomorphic. Then
$\widetilde\nabla_{\hat\xi}\Phi$ is a well defined (i.e. does not depend on how
we chose the lift $\hat\xi$),
smooth section of
$E^{corr}\ra Y$ whose restriction to each fiber $Y_s$ is holomorphic.
Now if $\widetilde\nabla_{\hat\xi}\Phi$ happens to be $L^2$ along each fiber
$Y_s$, then $(\widetilde\nabla_{\hat\xi}\Phi)^{\check{ }}$ yields a section of
$H^{corr}\ra S$. This gives the idea how to try
to define the quantum connection.
The elements $\varphi\in\Gamma^\infty$ should have the properties:
$\hat\varphi$ is a  smooth section of $E^{corr}$, holomorphic and $L^2$
along each $Y_s$ and for any
smooth vector field $\xi$ of type $(0,1)$ on $S$,
$(\widetilde\nabla_{\hat\xi}\hat\varphi)$ should have the same properties.
Then
$$
\nabla_\xi\varphi:=(\widetilde\nabla_{\hat\xi}\hat\varphi)^{\check{ }}.\tag1.2.5
$$
would define the quantum connection for $(0,1)$ type vector fields.

The quantum connection is supposed to be a Hermitian connection, i.e. for
a $(1,0)$ vector field $\xi$ and (appropriate) section $\varphi$ of
$H^{corr}\ra S$, 
$\nabla_\xi\varphi$ should  be that section $\psi$ of $H^{corr}\ra S$, which
corresponds to the pointwise continuous linear functional
$\theta\mt \xi h(\varphi,\theta)-
h(\varphi,\nabla_{\overline\xi}\theta)$, $\theta\in\Gamma^\infty$.

Finally if $\xi$ is any smooth vector field on $Y$ and
$\varphi\in\Gamma^\infty$,
$$
\nabla_\xi\varphi:=\nabla_{\xi^{0,1}}\varphi+\nabla_{\xi^{1,0}}\varphi.
$$
For more details and precise statements see [LSz3, Sect. 6-9].

\subhead 1.3. Symmetric spaces 
\endsubhead
%\medskip
Let $(M^m, g)$ be an $m$-dimensional,  compact,
 irreducible, simply connected, Riemannian symmetric space.
Then    $M$ is  isometric to $U/K$, where  $U$ is
a compact, connected, 
simply connected, semisimple
Lie group   and 
$K$ is the fixed point set (automatically connected)
of a  nontrivial involution $\theta:U\ra U$. The metric on $U/K$ is
induced from a biinvariant metric on $U$ ([He1]). 
 Furthermore either $U$ is simple or has the form $U=G\times G$ where
 $G$ is simple, $\theta(g_1,g_2)=(g_2,g_1)$
 and $K$ is the diagonal in $G\times G$. In the latter case
 $M$ is isometric to $G$ equipped with a biinvariant metric.

Let $\fu$ be the Lie algebra of $U$, $\fu_\bC$ its complexification and
$U_\bC$ the simply connected
complex Lie group with Lie algebra $\fu_\bC$. Since $U$ is compact, 
the canonical
Lie algebra embedding   $\iota_*:\fu\ra \fu_\bC$ yields an embedding
  $\iota:U\ra U_\bC$. 

 As a smooth manifold 
  $U_\bC$   naturally identifies with the tangent bundle $TU$.  The
complex structure on $TU$ obtained using this diffeomorphism will be the
adapted complex structure of a biinvariant metric on $U$
 (see [Sz2, Prop.3.5]). This is the complex structure that corresponds to
 the parameter $i$ from Sect. 1.1.

$\theta$ induces a Lie algebra involution $\theta_*:\fu\ra\fu$.
 Then $\fu=\fk+\fp_*$, where
$\fk=\{X\in\fu : \theta_*(X)=X\}$ and $\fp_*=\{X\in\fu : \theta_*(X)=-X\}$.
Here 
 $\fk$ is  the Lie algebra of $K$ and $\fp_*$ can be identified with 
$T_{[K]}M$. 

Let $\fp_0=i\fp_*$, $\fg_0=\fk+\fp_0$ and denote by $G_0$ the analytic subgroup
of $U_\bC$  with Lie algebra $\fg_0$. Then $G_0$ is closed in $U_\bC$ and
$K\subset G_0$. Let $\theta_\bC$ be the holomorphic extension of $\theta$ to
$U_\bC$. Then $\left.\theta_\bC\right|_{G_0}$ is a Cartan involution on $G_0$
with fixed point set $K$. The corresponding symmetric space $X=G_0/K$ is the
noncompact dual of $U/K$.

Let $\fa_*\subset\fp_*$
be a maximal Abelian subspace. Its dimension $r:=\dim\fa_*$ is the rank of $M$.
Let  $\fa_0:=i\fa_*$ and $\fh_0\subset\fg_0$ be  
a maximal Abelian subalgebra 
containing $\fa_0$. The complexification of $\fh_0$ (resp. of  $\fa_0$)
is $\fh$ (resp. $\fa$).
Let $\Delta$ be the set of nonzero roots corresponding to $(\fu_\bC,\fh)$ and
$\Sigma$  the  set of restricted roots.

Let $\fh_{\fk_0}=\fh_0\cap\fk$ and $\fh_\bR=\fa_0+i\fh_{\fk_0}$. The roots are real
valued on $\fh_\bR$. Choose a compatible ordering in the dual spaces of
$\fa_0$ and $\fh_\bR$. This yields an ordering of $\Delta$ and $\Sigma$.
Let
 $\rho_\Delta$ be half the sum of the positive roots and
 $\rho$ its restriction to $\fa$, i.e. 
$\rho=(1/2)\sum\limits_{\alpha\in\Sigma^+}m_\alpha\alpha$,
where $m_\alpha$ is
the multiplicity of $\alpha$. 
 $\fa_+\subset\fa_0$  denotes the positive Weyl chamber
$$
\fa_+:=\{H\in\fa_0 : \alpha(H)>0, \forall \alpha\in\Sigma^+   \}.
$$
The classification of compact, irreducible Riemannian
symmetric spaces shows that the restricted root system
together with the multiplicity function determines the symmetric space uniquely
(see [He1]). In particular we have
\proclaim{Proposition 1.3.1}
A compact, simply connected Riemannian symmetric space 
$M$ is isometric to a compact, simply connected Lie group equipped with a
biinvariant metric if and only if $\Sigma$ is a reduced root system and
each $m_\alpha$ is equal to $2$.
\endproclaim
(See [L, Theorem 4.4, p.82]).
We intend to use this characterization of Lie groups to prove Theorem 0.1.

\head 2. Curvature calculations
\endhead

\subhead 2.1. Flatness and projective flatness
\endsubhead

Consider a compact, simply connected, irreducible Riemannian symmetric space
$(M^m, g)$,  given in the form $M=U/K$ as in Sect. 1.3 and $H^{corr}\ra S$
 the corresponding field of quantum Hilbert spaces ($S$ being the complex
 upper half plane).

 $U$ acts on $(N,J(i))$ by biholomorphisms and this action 
 induces a representation $\hat \pi$ on $\Cal O(N, J(i))$, by the formula
$av=(a^{-1})^*v$ (pull back by $a^{-1}$), where $a\in U$, 
$v\in \Cal O(N, J(i))$. 
The same formula defines a unitary representation $\pi$ on $L^2(M)$. 
The restrictions $V_\chi|_M$ of the isotypical subspaces of   $\hat \pi$  are
precisely the isotypical subspaces of $\pi$ and the latter are well known to be
finite dimensional.
Since $M$ is a  maximal dimensional, totally real submanifold 
in $N$, we get that $V_\chi$
  are also finite dimensional.
  
The restrictions of $\hat\pi$ to the isotypical subspaces
 $V_\chi$ (or equivalently the restrictions
 of $\pi$ to $V_\chi|_M$) are irreducible, they are precisely  the 
irreducible $K-$spherical  representations of $U$ 
([He2, Chap. V, Theorem 4.3]).
  Therefore from now on we use
the spherical representations $\delta$ themselves  instead of their
character $\chi$, to label the objects (unlike in [L-Sz3]),
for example $V_\delta$
will replace $V_{\chi}$.

Flatness of the field $H^{corr}\ra S$ can be understood in terms of certain
Toeplitz operators $P_\delta(s)$ on $V_\delta$. They are $U$-equivariant, whence
according to Schur's lemma, have the form $P_\delta(s)=p_\delta(s)Id_{V_\delta}$ with
an appropriate function $p_\delta$.
 $H^{corr}\ra S$ is flat (resp. 
projectively flat) if and only if 
 $\bar\partial\partial\log p_\delta(s)=0$ for all $\delta$ (resp.  
$\bar\partial\partial\log p_\delta(s)$ is independent of $\delta$), 
see [L-Sz3, Theorem 9.2.1].

 In our situation according to [L-Sz3, Lemma 11.2.1]
 $p_\delta(s)$ depends only
on $\tau=\text{Im}s$ and has the specific form
$$
p_\delta(s)=Cc_\delta \tau^{-m/2}q_\delta(\tau),\tag2.1.1
$$
where $m$ is 
the dimension of the space $M$, $C$ is some constant, $c_\delta$
 a constant  for each representation $\delta$ and
 $q_\delta$  an appropriate function (see (2.2.6)  for the precise form).
As one  easily sees, a factor like $C\tau^{-m/2}$ that depends only on
$\tau=\text{Im}s$ but not on $\delta$ does not affect the condition for
projective flatness.
In our case, in light of (2.1.1), 
 the above mentioned characterization of (projective) flatness takes the form.
 \proclaim{Theorem 2.1.1}
\item{(a)} $H^{corr}\ra S$ is flat iff for each $\delta$, $\log(p_\delta(s))$
is harmonic.
%  $A_\delta>0$, $B_\delta$ with
%  $q_\delta(\tau)=\tau^{\frac{m}{2}}A_\delta e^{B_\delta\tau}$.
\item{(b)}
$H^{corr}\ra S$ is projectively flat iff for each $\delta$ there exist 
  constants $A_\delta>0$, $B_\delta$ with
  $q_\delta(\tau)=A_\delta e^{B_\delta\tau}q_{\delta_0},$ where $\delta_0$ denotes
  the trivial representation.
\endproclaim
Since we cannot compute $q_\delta$ explicitly, we cannot
 check directly whether condition (b) in Theorem 2.1.1 holds or not.
 Therefore we shall apply the
following strategy to prove Theorem 0.1.
We shall investigate the
asymptotic behavior of $q_\delta(\tau)$ as $\tau$ tends to $0$ and to infinity.
From the $\tau\ra 0$ asymptotics we shall determine the values of
$A_\delta$ and $B_\delta$ dictated by condition (b) in Theorem 2.1.1.
Then do the same as $\tau\ra\infty$ and obtain possible different values
for $A_\delta$ and $B_\delta$. If the values for $A_\delta$ or $B_\delta$
do not match as $\tau\ra 0$ and as $\tau\ra\infty$,
we can conclude that the Hilbert field is not projectively flat.

It turns out, that  $B_\delta$ does not help in determining the projective
flatness of $H^{corr}\ra S$,
 for all rank-1 symmetric spaces the two asymptotics give the same
 value for $B_\delta$ (see Remark 1, after Theorem 5.4).
 Theorem 0.1 is proved by showing that
 the $\tau\ra 0$ asymptotics yields
  $A_\delta=1$ for all $\delta$ (see Theorem 4.2.2),
  on the other hand
  the $\tau\ra\infty$ asymptotic shows that
 if the coefficient $A_\delta$ is independent of $\delta$, 
the restricted root system of $M$ must be reduced and all multiplicities
of the roots are equal to two (see Sect. 5 and 6). But these properties
characterize compact Lie groups among compact Riemannian symmetric spaces
(see [L]) and Theorem 0.1 will follow.

\subhead 2.2. The function $q_\delta(\tau)$
\endsubhead

Now to implement the  plan in Sect. 2.1,
we need to recall first of all the precise
form of $p_\delta(s)$ (see [L-Sz3, Sect. 12.1], $\tau=$Im$s$).  
$$
p_\delta(s)=\frac{c_\delta }{\tau^{m/2}}
\int\limits_{\frak p_*}\int\limits_K 
e^{-\frac{|\zeta|^2}{\tau}}\chi_\delta (k\exp (-2i\zeta))
\sqrt{\eta(\zeta)}
dk \,d\zeta,\tag2.2.1
$$
where $c_\delta$ is independent of $s$,
$dk$ is normalized Haar measure on $K$, $d\zeta$ is the  Lebesgue measure 
on $\frak p_*$ induced by the metric, $\chi_\delta$ the character of $\delta$
and  
 $$
\eta(\zeta)=\det \left(\left.\frac{\sin 2\ad\zeta}{\ad\zeta}
\right|_{\bC\otimes\frak p_*}\right).\tag2.2.2
$$
The function $f_\delta(g)=\int_K\chi_\delta(k g^{-1})dk$,
$g\in U$ is known as the  
$K-$spherical function ([Ha1,Ha2]), corresponding to the representation
$\delta$, see 
[He2, Theorem 4.2, p.417]. We denote by the same letter
the holomorphic extension of $f_\delta$
to the complexified group $U_\bC$.
 Thus we can rewrite (2.2.1) as an integral over $\fp_0$ and we get
$$
p_\delta(s)=\frac{c_\delta }{\tau^{m/2}}
\int\limits_{\frak p_0} 
e^{-\frac{|H|^2}{\tau}}f_\delta (\exp (2H))
\sqrt{\eta(-iH)} \,dH.\tag2.2.3
$$

Every restricted root  $\alpha\in\Sigma$ is real valued on $\fa_0$. 
Furthermore    the operator $ad_H^2$, 
  $H\in\fp_0$, is  symmetric,
has zero eigenvalue with multiplicity $r=\dim\fa_0$
and $\alpha(H)^2$ with multiplicity $m_\alpha$. Thus 
from (2.2.2) and the identity $\sin i2z/iz=\sh2z/z$ we get 
$$
\eta(-iH)=2^r\prod\limits_{\alpha\in\Sigma^+}\left(
\frac{\sh(2\alpha(H))}{\alpha(H)}\right)^{m_\alpha}.\tag2.2.4
$$

Let $C(\fa_0):=\{k\in K : Ad(k)\zeta=\zeta,\  \forall \zeta\in\fa_0\}$ 
be the centralizer of $\fa_0$ in $K$. Recall the following integral formula
for  the generalized polar coordinate map 
$$
\Phi:(K/C(\fa_0))\times\fa_0\ra \fp_0,\quad
\Phi(kC(\fa_0),H):=Ad(k)H,
$$

\proclaim{Theorem 2.2.1} Let $f\in L^1(\fp_0)$ be an $Ad(K)$ invariant
function. Then
$$
\int\limits_{\frak p_0}f(H)dH=c\int\limits_{\frak a_+}f(H)
\prod\limits_{\alpha\in\Sigma^+}\alpha(H)^{m_\alpha}dH,
$$
where $c$ is some constant, independent of $f$ and
$\frak a_+\subset\fa_0$  denotes the positive Weyl chamber.
\endproclaim
(See [He2, Theorem 5.17, p.195].)

From [L-Sz4, Prop. 2.1 and Prop. 2.2] we know that $f_\delta\circ\exp$ and $\eta$
are $Ad_K$ invariant on $\fp_0$. Thus Theorem 2.2.1, (2.2.3) and (2.2.4)
yields the
following formula.% for $p_\delta$.
$$
p_\delta(s)=2^rc_\delta c\tau^{-m/2}
\int\limits_{\frak a_+} 
e^{-\frac{|H|^2}{\tau}}f_\delta (\exp (2H))
\prod\limits_{\alpha\in\Sigma^+}\left(\alpha(H)
\sh(2\alpha(H))\right)^{\frac{m_\alpha}{2}}
dH.\tag2.2.5
$$
In the special case when
$M$ is isometric to a compact Lie group $G$, let $U=G\times G$ and $K$ 
the diagonal in $U$. Then  the $K-$spherical functions will have the form
$f_\delta=\chi_\delta/d(\delta)$, where $\delta$ is an irreducible representation
of $G$, $\chi_\delta$  its character and $d(\delta)$ denotes its dimension
([He2, p.407]). Thus $f_\delta$ is given by the Weyl character formula and
since  all $m_\alpha=2$ the terms $\sh(2\alpha(H))$ cancel out the Weyl
denominator and we end up essentially integrating the product of a Gaussian
and a harmonic polynomial. This yields that $\log(p_\delta(s))=c_1+c_2$Im$s$,
 a harmonic function (see [LSz3, Theorem 11.3.1]). Thus in light of
 Theorem 2.1.1(a), the field
 $H^{corr}\ra S$ will be flat.

To treat the  other symmetric spaces, we introduce the  essential part of
$p_\delta$ as a
function of $\tau>0$:  %by the  integral part of $p_\delta$:
$$
q_\delta(\tau):=\int\limits_{\frak a_+} 
e^{-\frac{|H|^2}{\tau}}f_\delta (\exp (2H))
\prod\limits_{\alpha\in\Sigma^+}\left(\alpha(H)
\sh(2\alpha(H))\right)^{\frac{m_\alpha}{2}}
dH.\tag2.2.6
$$
%Thus $p_\delta(s)=\frac{2^rc_\delta Vol(K/M) }{(\text{Im }s)^{m/2}}
%q_\delta(\text{Im}s)$.
 %In light
%of our earlier characterization of the projective flatness of $H^{corr}\ra S$
%in terms of $\bar\partial\partial\log p_\chi$,
% from (2.5) and (2.6) we obtain.

%\proclaim{Theorem 2.2} $H^{corr}\ra S$ is projectively flat iff
%$(\log q_\delta(x))''$ does not depend on the
%irreducible $K-$spherical representation $\delta$ of $U$, i.e.
%for every  such representation
%there exist constants $A_\delta, B_\delta\in\bR$ such that
%$$
%q_\delta(x)=A_\delta e^{B_\delta x}q_{\delta_0}(x),
%$$
%where $\delta_0$ is the trivial representation.
%\endproclaim

\head 3.  Spherical functions
\endhead
\medskip
In order to be able to handle the integral in (2.2.6),
we shall need another description of spherical functions.
Let $\delta:U\ra GL(V)$ be an irreducible $K-$spherical 
representation.  We can endow $V$ with a scalar product $\langle.,.\rangle$
 that makes $\delta$ unitary.
Let $v_K\in V$ be a $K-$fixed vector with unit length. 
Then  the spherical function  $f_\delta$ corresponding to $\delta$ is
([He2, Theorem 3.7, p.414])
$$
f_\delta(g):=\langle\delta(g)v_K,v_K\rangle,\quad g\in U.
$$
Since $\delta$ extends holomorphically to the complexified group
$U_\bC$, the same formula
yields  the holomorphic extension of $f_\delta$  to 
$U_\bC$.

We would like to obtain some formula for the function $f_\delta\circ\exp$,
occurring in (2.2.6),
when we restrict it  to the Cartan subalgebra $\fh$ in $\fu_\bC$.

Let
$\Lambda(\delta)$ be the set of weights of $\delta$
and for a weight $\mu$,  $W_\mu$  the corresponding weight space.

 The weight spaces give an orthogonal direct decomposition of $V$, thus 
$$
v_K=\sum\limits_{\mu\in\Lambda(\delta)}w_\mu,\quad w_\mu\in W_\mu,
$$
where 
$\|v_K\|=1$ implies $\sum\|w_\mu\|^2=1$. 

Let $H\in\fh$. Then (cf. [V])
$$
f_\delta(\exp 2H)=\langle\exp(\delta_*2H)v_K,v_K\rangle=
\sum\limits_{\mu\in\Lambda(\delta)}e^{2\mu(H)}\langle w_\mu,w_\mu\rangle.\tag3.1
$$

Later on  we shall need to figure out which term in (3.1) has the
dominating contribution
when (3.1) is plugged into the formula (2.2.6) of $q_\delta$. It is no surprise
that the term corresponding to the highest weight will play this role.
 Theorem 5.2 will give  the precise answer. That theorem will be based on
 Theorem 5.1, a general result on asymptotics of integrals of the form
 (2.2.6), where the function $f_\delta$ is replaced by an exponential of a linear
 function, like the terms in (3.1).
  The result of Theorem 5.1 shall explain why we need Proposition 3.1. 
   
Let $\lambda$ be the highest weight of $\delta$. Then $\dim W_\lambda=1$. Let
$v_\lambda\in W_\lambda$ with $\|v_\lambda\|=1$. Thus
$w_\lambda=a_\lambda v_\lambda$ with
$a_\lambda=\langle v_K,v_\lambda\rangle.$
From the first formula in [He2, p.538] we know that $a_\lambda\not=0$ and 
$$
\langle w_\lambda,w_\lambda\rangle=|a_\lambda|^2=\bold c(-i\lambda-i\rho),\tag3.2
$$
where  $\rho=\left.\rho_\Delta\right|_{\fa_0}$
is half the sum of the positive restricted 
roots with multiplicity, $X=G/K$ the noncompact dual symmetric space
and {\bf c} is the corresponding Harish-Chandra's
{\bf c}-function of $G$ ([Ha1, Ha2], [He2, (8), p.538]).

%$\fa_0:=i\fa_*$ is maximal Abelian in $\fp_0:=\fp_*$ and 
%the weights are real valued
%on $\fa_0$. %The Killing form induces a Riemannian metric on $U$.

\proclaim{Proposition 3.1}
Let $\mu\in \Lambda(\delta)$, $\mu\not=\lambda$. Then
$$
\|\left.(\mu+\rho_\Delta\right)|_{\fa_0}\|<\|
\left.(\lambda+\rho_\Delta\right)|_{\fa_0}\|.
$$
\endproclaim
\demo{Proof}
We follow the steps of the proof of [He2, Theorem. 1.3, p.498],
that is the same statement
without taking  restrictions to $\fa_0$.
First we show that
$$
\left.(\lambda-\mu)\right|_{\fa_0}\not\equiv0.\tag3.3
$$
Since $\lambda$ is the highest weight of a $K-$spherical representation,
the Cartan-Helgason theorem ([He2, Theorem 4.1 (1), p.535]) implies 
$$\left.\lambda\right|_{i\fh_{\fk_0}}\equiv0.$$
Thus if (3.3) does not hold, we  would get
$\langle\lambda-\mu,\lambda\rangle=0$ and then
$$
\langle\mu,\mu\rangle=
\langle\mu-\lambda,\mu-\lambda\rangle+
\langle\lambda,\lambda\rangle>\langle\lambda,\lambda\rangle,\tag3.4
$$
since $\mu\not=\lambda$.  
But (3.4) contradicts  the fact that for all weights $\mu$, 
$\|\mu\|\le\|\lambda\|$ (see [He2, Theorem 1.3 (7), p.498]) and
so (3.3) is proved.

We need to show 
$$
C:=\|\lambda+\rho\|^2-\|\left.\mu\right|_{\fa_0}+\rho\|^2>0.
$$
But
$$
C=\|\lambda\|^2-\|\left.\mu\right|_{\fa_0}\|^2+
2\langle\lambda-\left.\mu\right|_{\fa_0},\rho\rangle\ge
\|\lambda\|^2-\|\mu\|^2+
2\langle\lambda-\left.\mu\right|_{\fa_0},\rho\rangle.\tag3.5
$$
And since  $\|\lambda\|\ge\|\mu\|$, it suffices to 
show that the last term in (3.5) is positive.

 Let $\alpha_1,\dots,\alpha_l$ be a basis of the roots,
 compatible with $\Sigma$, i.e. for
 $1\le j\le r$ $\left.\alpha_j\right|_{\fa_0}\in\Sigma^+$
 forming a basis of
 $\Sigma$.
Since $\mu$ is a weight,
 $\exists n_j\in\Bbb Z_+$ with $\mu=\lambda-\sum\limits_1^ln_j\alpha_j$.
Now (3.3) implies that
$\exists j$ with $1\le j\le r$ and $n_j>0$.
  Proposition 3.2  shows that 
 $\langle\left.\alpha_j\right|_{\fa_0},\rho\rangle>0$ for $1\le j\le r$. Hence
$$
\langle\left.\lambda-\mu\right|_{\fa_0},\rho\rangle=\sum\limits_1^rn_j
\langle\left.\alpha_j\right|_{\fa_0},\rho\rangle>0,
$$
thus indeed $C>0$.
\qed
\enddemo
\proclaim{Proposition 3.2} Let $\alpha_1,\dots,\alpha_r\in\Sigma^+$
be a basis of the
restricted roots $\Sigma$ with multiplicities $m_{\alpha_j}$. Then
$$
\langle\rho,\alpha_j\rangle=(m_{\alpha_j}/2+m_{2\alpha_j})
\langle\alpha_j,\alpha_j\rangle,
\quad j=1\dots,r\tag3.6
$$ 
where $m_{2\alpha_j}$ is meant to be zero if $2\alpha_j$ is not a root.
\endproclaim
\demo{Proof}
Let $\Sigma^+_{j}=\Sigma^+\setminus\{\alpha_j,2\alpha_j\}$, 
$\rho_j=\frac1{2}\sum\limits_{\alpha\in \Sigma^+_j}m_\alpha\alpha$ and 
$S_{\alpha_j}$ the reflection on $\fa_0$, corresponding to $\alpha_j$.  
As is well known ([He1, ChVII, Sect. 3, Lemma 2.21])
$S_{\alpha_j}$ permutes the elements of $\Sigma^+_j$, hence 
$S_{\alpha_j}\rho_j=\rho_j$.
From their definitions we get
$$
\rho=\rho_j+\frac{m_{\alpha_j}\alpha_j+m_{2\alpha_j}2\alpha_j}{2}.
$$
 Thus 
$$
S_{\alpha_j}\rho=\rho-m_{\alpha_j}\alpha_j-m_{2\alpha_j}2\alpha_j.
$$
Since $S_{\alpha_j}$ is an orthogonal transformation, we obtain
$$
\langle\rho,\alpha_j\rangle=\langle S_{\alpha_j}\rho,S_{\alpha_j}\alpha_j\rangle=
\langle\rho-m_{\alpha_j}\alpha_j-m_{2\alpha_j}2\alpha_j,-\alpha_j\rangle
$$
and (3.6) follows.
\qed
\enddemo
\head 4. $\tau\ra 0$ asymptotics
\endhead

\subhead 4.1. A multivariable Watson lemma
\endsubhead

\proclaim{Proposition 4.1.1}
Let  $0<\tau$, $0<h$, $D\subset\bR^n$  be a domain that is
 a homogeneous cone
($\xi\in D, 0<r$ implies $r\xi\in D$), 
$G:=D\cap S^{n-1}$  (where $S^{n-1}$ is the unit sphere in $\bR^n$)
  and $Q$ an $h$-homogeneous 
 (for all $\xi\in D, 0<r$, $Q(r\xi)=r^hQ(\xi)$ ) 
continuous function defined on $\overline D$.  
Then
$$
\int\limits_De^{\frac{-\|H\|^2}{\tau}}Q(H)dH=\frac{\Gamma(\frac{n+h}{2})}{2}
\left(\int\limits_GQ(\xi)d\xi\right) \tau^{\frac{n+h}{2}},
$$
where $\Gamma$ denotes the usual gamma function.
\endproclaim
\demo{Proof}
Using polar coordinates and the homogeneity of $Q$ we get
$$
\int\limits_De^{\frac{-\|H\|^2}{\tau}}Q(H)dH=
\int\limits_0^\infty\int\limits_Ge^{\frac{-r^2}{\tau}}r^{n+h-1}Q(\xi)d\xi dr.
$$
Substituting $r=\sqrt{\tau t}$ yields the formula.
\qed
\enddemo

\proclaim{Proposition 4.1.2} Let $\delta, \tau_0>0$,  
$D\subset\bR^n$  be a domain,  $D_\delta:=D\cap\{\|H\|\ge\delta\}$ and $g$  a
Lebesgue measurable function on $D$ with 
$$
C:=\int\limits_De^{-\frac{\|H\|^2}{\tau_0}}|g(H)|dH<\infty.
$$ 
 Then for every
$0<\tau<\tau_0$
$$
\int\limits_{D_\delta}e^{\frac{-\|H\|^2}{\tau}}|g(H)|dH\le
Ce^{\delta^2(\frac1{\tau_0}-\frac1{\tau})} .
$$
\endproclaim
\demo{Proof}
Let $\delta\le\|H\|$. Then 
$$
e^{\|H\|^2(\frac1{\tau_0}-\frac1{\tau})}\le 
e^{\delta^2(\frac1{\tau_0}-\frac1{\tau})}.
$$ 
Thus
$$
\aligned
\int\limits_{D_\delta}e^{\frac{-\|H\|^2}{\tau}}|g(H)|dH&=
\int\limits_{D_\delta}e^{\frac{-\|H\|^2}{\tau_0}}|g(H)|e^{\|H\|^2(\frac1{\tau_0}-
\frac1{\tau})}dH\le\\
&\le\int\limits_{D_\delta}e^{\frac{-\|H\|^2}{\tau_0}}|g(H)|
e^{\delta^2(\frac1{\tau_0}-\frac1{\tau})}dH.
\qed
\endaligned
$$
\enddemo

\proclaim{Theorem 4.1.3} Let $0<a\le\infty$,  $G$ be  a
 domain in  $S^{n-1}$ (unit sphere), $0<d$,
$$
G_a:=\{r\xi : 0<r<a, \xi\in G\}
$$ 
 and $Q$  a 
$d-$homogeneous continuous function defined on $\overline G_a$.
Suppose $f\in C(G_a)$
that is  $ C^\infty$ in a neighborhood of the origin.  
Assume that for some $0<\tau_0$ the function  $e^{-\|H\|^2/\tau_0}Q(H)f(H)$ is in 
$L^1(G_a)$.
  For $0<\tau<\tau_0$  let $\Phi(\tau)$  be defined by
$$
\Phi(\tau)=\int\limits_{G_a}e^{-\frac{\|H\|^2}{\tau}}Q(H)f(H)dH.
$$
Then $\Phi$ admits an asymptotic series expansion around $0$:
$$
\Phi(\tau)\sim\sum\limits_{j=0}^\infty\frac{\Gamma(\frac{n+d+j}{2})}{2}
\int\limits_GQP_jd\xi\ \tau^{\frac{n+d+j}{2}},
\quad \tau\ra 0,
$$
where $P_j$ is the $j-th$ homogeneous polynomial term of the Taylor series of
$f$ around the origin.
\endproclaim
\demo{Proof}
We follow the scheme of the proof of  Watson's lemma in one variable (cf. [M]).
 Let $0<\delta\le a$ be so small that  $f$ is $C^\infty$ in a 
neighborhood of the ball $\overline{\Bbb B}_\delta^n(0)$.
Then $G_a\cap\Bbb B_\delta^n(0)=G_\delta$ and with 
$h(\tau,H)=e^{-\|H\|^2/\tau}Q(H)f(H)$
$$
\Phi(\tau)=
\int\limits_{G_a\cap\{\|H\|\ge\delta\}}h(\tau,H)dH + 
\int\limits_{G_\delta}h(\tau,H)dH
=:\Phi_1(\tau)+\Phi_2(\tau).
$$
With $g(H)=Q(H)f(H)$ and $C=\int\limits_{G_a}e^{-\frac{\|H\|^2}{\tau_0}}|Q(H)f(H)|dH$, 
Proposition 4.1.2 implies
$$
|\Phi_1(\tau)|\le C e^{\frac{\delta^2}{\tau_0}} e^{-\frac{\delta^2}{\tau}}=o(\tau^n),
\quad \tau\ra 0,
$$
for all $n\in\Bbb N$.
The Taylor formula with remainder term yields 
$$
f(H)=\sum\limits_{j=0}^NP_j(H)+f_N(H),\quad \|H\|\le\delta,\quad 
|f_N(H)|\le C_N\|H\|^{N+1},\tag4.1.1
$$
where $P_j$ is a $j-$homogeneous polynomial and $C_N$
 an appropriate constant. Thus
$$
\Phi_2(\tau)=\sum\limits_{j=0}^N\ \int\limits_{G_\delta}e^{-\frac{\|H\|^2}{\tau}}Q(H)
P_j(H)dH+\int\limits_{G_\delta}e^{-\frac{\|H\|^2}{\tau}}Q(H)f_N(H)dH
$$
and
$$\aligned
\int\limits_{G_\delta}e^{-\frac{\|H\|^2}{\tau}}Q(H)P_j(H)dH=
\int\limits_{G_\infty}e^{-\frac{\|H\|^2}{\tau}}Q(H)P_j(H)dH-\\
\int\limits_{G_\infty\cap\{\|H\|\ge\delta\}}e^{-\frac{\|H\|^2}{\tau}}Q(H)P_j(H)dH.
\endaligned
$$
In light of Proposition 4.1.1  the first integral on the right hand side is
 equal to
$$
\frac{\Gamma(\frac{n+d+j}{2})}{2}
\left(\int\limits_GQP_jd\xi\right)\tau^{\frac{n+d+j}{2}},
$$
and Proposition 4.1.2
yields with $g=QP_j$, that the second integral is $o(\tau^n)$
for all  $n\in\Bbb N$.
Homogeneity of $Q$ implies $|Q(H)|\le K\|H\|^d$ with some $K>0$.
Then  from (4.1.1) and Proposition 4.1.1 we get 
$$\aligned
\left|\int\limits_{G_\delta}e^{-\frac{\|H\|^2}{\tau}}Q(H)f_N(H)dH\right|\le C_NK
\int\limits_{G_\infty}e^{-\frac{\|H\|^2}{\tau}}\|H\|^{N+d+1}dH=\\
 =Vol(G)C_NK\frac{\Gamma(\frac{n+d+N+1}{2})}{2}\tau^{\frac{n+d+N+1}{2}},
\endaligned
$$
finishing the proof of the theorem.
\qed
\enddemo

\subhead 4.2. Determining $A_\delta$ and $B_\delta$ from $\tau\ra0$
\endsubhead

Let us get back to the symmetric space situation.
Suppose $(M^m=U/K,g)$ is a compact, simply connected,  irreducible, 
Riemannian symmetric space as in Sect. 1. As before, $m$ is the dimension
of $M$.
Let $\delta$ be an irreducible $K-$spherical
representation and $f_\delta$  the corresponding spherical function.
Then
$$
f_\delta(\exp(2H))=1+R_1(H)+R_2(H)+\dots,\quad H\in\fa_0,\tag4.2.1
$$
where $R_j$ is the $j-$th homogeneous polynomial term of the Taylor series.
Since $f_\delta\circ\exp$ is Ad$_K$ invariant on $\fp_0$
(see [L-Sz4, Proposition 2.1]),
it is Weyl group invariant on $\fa_0$. Therefore every $R_j$ is Weyl
group invariant as well.
Since $M$ is irreducible, the Weyl group acts irreducible on $\fa_0$, thus
$R_1\equiv0$ and 
$R_2$  must be of the form 
$$
R_2(H)=b_\delta\|H\|^2,\tag4.2.2
$$
with some  $b_\delta\in\bR$. (4.2.2) is true because up to a
constant scalar, $\|H\|^2$ is the only  Weyl group invariant
quadratic polynomial on $\fa_0$.
One can see this either as a corollary of Schur's lemma,
or as a corollary of 
  Chevalley's theorem (see [Hu, Sect. 3.5,
3.7]). For the trivial representation $\delta_0$, $f_{\delta_0}\equiv1$ and
$b_{\delta_0}=0$.

\proclaim{Proposition 4.2.1} Assume that the rank of $M$ is $1$ and $\lambda$
is the highest weight of $\delta$. Then
$$
b_\delta=\frac{2(\|\lambda+\rho\|^2-\|\rho\|^2)}{m}.
$$
\endproclaim
\demo{Proof} If $\Sigma$ is nonreduced, $\Sigma^+=\{\beta, \beta/2\}$
 and  $\Sigma^+=\{\beta\}$ in the reduced case.
The corresponding multiplicities are $m_\beta$ and $m_{\beta/2}$, where
our convention is that the latter is zero when $\Sigma$ is reduced.
Let $H_0\in \fa_+$ with $\|H_0\|=1$. Then $\beta(H_0)=\|\beta\|$.
Recall that Gauss' hypergeometric functions 
are given by
$$
F(a,b,c,z):=1+\frac{ab}{c}z+\dots
+\frac{a(a+1)\dots(a+k-1)b(b+1)\dots(b+k-1)}{k!c(c+1)\dots(c+k-1)}
z^k+\dots
$$
where $a,b,c\in\Bbb C$, $c\not\in \Bbb Z_{-}=\{0,-1,-2,\ldots\}$. The series
 converges at least in the unit disk. 
If $n\in\Bbb Z_+$, $b=-n$, $A\in\Bbb C\setminus\Bbb Z_-$,  and 
$a=A+n$,  then  $F$  is a
polynomial (in $z$)  of degree $n$.

 According to [He2, Theorem 4.1(ii), p. 535
 and Sect. 3, p. 542] the highest weight of $\delta$
 has the form $\lambda=n_\delta\beta$, where $n_\delta\in\Bbb Z_+$.
Let 
$$
a_\delta:=\frac1{2}m_{\beta/2}+m_\beta+n_\delta,\quad
c_\delta:=\frac{m_{\beta/2}+m_\beta+1}{2}=\frac{m}{2}.
$$
Denote by $F_\delta$ the hypergeometric function (polynomial in this
case),
 corresponding to these parameters
$$
F_\delta(x)=F(a_\delta,-n_\delta,c_\delta,x).
$$
According to 
\cite{He2, formula (25), p.543}, the spherical function $f_\delta$ 
can be expressed as 
$$
f_\delta(\exp(2H))=F_\delta(-\sh^2(\beta(H))),\quad H\in\fa_0.
$$
Hence
$$
f_\delta(\exp(2H))=1+\frac{a_\delta n_\delta}{c_\delta}\|\beta\|^2\|H\|^2+o(\|H\|^2).
$$
Thus $b_\delta=\frac{a_\delta n_\delta}{c_\delta}\|\beta\|^2$. Now $\rho=
\frac1{2}(m_{\beta/2}\beta/2+m_\beta\beta)$, hence
$$
a_\delta n_\delta\|\beta\|^2=2\langle\rho,\lambda\rangle+\|\lambda\|^2,
$$
and our statement follows.
\qed
\enddemo
The $\tau\ra0$ asymptotics yields the following values for $A_\delta$,
$B_\delta$ in Theorem 2.1.1 (b).
\proclaim{Theorem 4.2.2} Suppose the corrected field of quantum Hilbert spaces
$H^{corr}\ra S$ is projectively flat. Then for every irreducible $K-$spherical
representation $\delta$, 
%$$
%q_\delta(x)=e^{\frac{m}{2}b_\delta x}q_{\delta_0}(x)
%$$
%with an appropriate constant $b_\delta$, where $\delta_0$ denotes the trivial 
%representation.
$$
A_\delta=1,\quad B_\delta=\frac{m}{2}b_\delta.
$$
\endproclaim

\demo{Proof} 
Easy calculation shows that
$$
F(H):=\prod\limits_{\alpha\in\Sigma^+}\left(
\frac{\sh(2\alpha(H))}{\alpha(H)}\right)^{\frac{m_\alpha}{2}}=1+\sum
\limits_{\alpha\in\Sigma^+}\frac{m_\alpha}{3}\alpha^2(H)+\dots\tag4.2.3
$$
From (4.2.2) and (4.2.3) we obtain that in the 
homogeneous polynomial series expansion of 
$$
f_\delta(\exp(2H))F(H)=1+P^\delta_2(H)+P^\delta_3(H)+\dots,
$$
the quadratic term is
$$
P^\delta_2(H)=b_\delta\|H\|^2+\sum
\limits_{\alpha\in\Sigma^+}\frac{m_\alpha}{3}\alpha^2(H)
=b_\delta\|H\|^2+P^{\delta_0}_2(H).\tag 4.2.4
$$
Now $Q(H):=\prod\limits_{\alpha\in\Sigma^+}\alpha(H)^{m_\alpha}$ is a homogeneous
polynomial of degree
$$
d=\sum\limits_{\alpha\in\Sigma^+}m_\alpha=m-r,
$$
where 
$r=\dim\fa_0$ is the rank of $M$.
Applying Theorem 4.1.3 with $f$, $Q$, $a=\infty$, $G=\fa_+\cap S^{r-1}$ 
 we obtain
$$
q_\delta(\tau)=\frac{\Gamma(\frac{m}{2})}{2}\int\limits_GQ(\xi)d\xi\
\tau^{\frac{m}{2}}+
\frac{\Gamma(\frac{m+2}{2})}{2}
\int\limits_GQ(\xi)P^\delta_2(\xi)d\xi\ \tau^{\frac{m+2}{2}}
+o(\tau^{\frac{m+2}{2}}).\tag4.2.5
$$
Since the restricted roots are positive on the Weyl chamber $\fa_+$, 
we get $\int\limits_GQ(\xi)d\xi>0$.
Now writing out (4.2.5) for both $\delta$ and the trivial representation
$\delta_0$, 
 comparing the  coefficients  of the $\tau^{\frac{m}{2}}$
 term in the asymptotic series and using
Theorem 2.1.1 (b) we obtain
$A_\delta=1$. Then comparing the coefficients  of the $\tau^{\frac{m+2}{2}}$
as well, we obtain
$$
 B_\delta\frac{\Gamma(\frac{m}{2})}{2}
 \int\limits_GQ(\xi)d\xi=\frac{\Gamma(\frac{m+2}{2})}{2}\int\limits_GQ(\xi)
 (P^\delta_2(\xi)-P^{\delta_0}_2(\xi)) d\xi.\tag4.2.6
$$
From (4.2.4) we get $P^\delta_2(\xi)-P^{\delta_0}_2(\xi)=b_\delta\|\xi\|^2=b_\delta$,
since
$G$ is part of
the unit sphere. Thus (4.2.6) yields $B_\delta=\frac{m}{2} b_\delta$.
\qed
\enddemo

\head 5. Asymptotics at infinity
\endhead
\medskip
 The following setting is motivated by the system of restricted roots of a 
compact Riemannian symmetric space.

Let $(Z,\langle.,.\rangle)$ be a Euclidean space of dimension $r$
 and $\Sigma^+\subset Z^*$ a finite set so that
$$
Z_+:=\{H\in Z\mid \alpha(H)>0, \forall\alpha\in\Sigma^+\}
$$
is nonempty. For each $\alpha\in\Sigma^+$, let  $m_\alpha>0$ be given and 
define
$$
m:=r+\sum\limits_{\alpha\in\Sigma^+} m_\alpha,\quad 
\rho:=\frac1{2}\sum\limits_{\alpha\in\Sigma^+} m_\alpha\alpha. 
$$
For a linear functional $l:Z\ra\bR$,    define $A_l\in Z$ by
$l(H)=<A_l,H>$, $H\in Z$. Then   $\langle l,L\rangle:=
\langle A_l,A_L\rangle$,  $l,L\in Z^*$, defines an inner product on $Z^*$.
Let $f:Z_+\ra\bR$ be any measurable function. Assuming the integral below is
finite, introduce the following function, 
defined for $\tau>0$.
$$
q(\tau,f)=\int\limits_{Z_+}e^{-\frac{\|H\|^2}{\tau}}f(H)\prod\limits_{\alpha\in\Sigma^+}
\left(\alpha(H)\sh(2\alpha(H))\right)^{\frac{m_\alpha}{2}}dH.\tag5.1
$$
%Thus $q_\delta(x)=q(x,f_\delta(\exp(2.)))$

With $\mu\in Z^*$, let $I_\mu(\tau):=q(\tau, e^{2\mu})$.
Even though it is impossible to calculate  precisely  this integral (except
in some special cases), it is possible to 
determine the order of its 
 magnitude as $\tau\ra\infty$, and that suffices for  our purposes.

\proclaim{Theorem 5.1} For any $\mu\in Z^*$
$$I_\mu(\tau)=
\cases
2^{r-m}\pi^{\frac{r}{2}}\prod\limits_{\alpha\in\Sigma^+}
\langle\mu+\rho,\alpha\rangle^{\frac{m_\alpha}{2}}\tau^{\frac{m}{2}}
e^{\tau\|\mu+\rho\|^2}
(1+o(1)),& 
A_{\mu+\rho}\in Z_+\\
\tau^{\frac{m}{2}}e^{\tau\|\mu+\rho\|^2}o(1),& A_{\mu+\rho}\in Z\setminus Z_+
\endcases
$$
as $\tau\ra\infty$.
\endproclaim

\demo{Proof}
Factoring out $e^{m_\alpha\alpha(H)}$ from the product for each
$\alpha\in\Sigma^+$, we get
$$
I_\mu(\tau)=2^{r-m}\int\limits_{Z_+}e^{-\frac{\|H\|^2}{\tau}+2\mu(H)+2\rho(H)}
\prod\limits_{\alpha\in\Sigma^+}\alpha(H)^{\frac{m_\alpha}{2}}
(1-e^{-4\alpha(H)})^{\frac{m_\alpha}{2}}dH
$$
Now
$$
-\|H\|^2/\tau+2\mu(H)+2\rho(H)=-\|H/\sqrt\tau-\sqrt\tau
A_{\mu+\rho}\|^2+\tau\|\mu+\rho\|^2.
$$
Thus
$$
I_\mu(\tau)=\frac{e^{\tau\|\mu+\rho\|^2}}{2^{m-r}}\int\limits_{Z_+}
e^{-\|H/\sqrt\tau-\sqrt\tau A_{\mu+\rho}\|^2}
\prod\limits_{\alpha\in\Sigma^+}\alpha(H)^{\frac{m_\alpha}{2}}
(1-e^{-4\alpha(H)})^{\frac{m_\alpha}{2}}dH
$$
Let $\Phi_\tau(Y)$ be the affine linear change of coordinates in $Z$
 defined by 
$$
\Phi_\tau(Y):=\sqrt\tau Y+\tau A_{\mu+\rho}.
$$ 
Then
$\det \Phi_\tau'=\tau^{\frac{r}{2}}$ and with  $H=\Phi_\tau(Y)$, 
$$
\alpha(H)=\alpha(\sqrt \tau Y+\tau A_{\mu+\rho})=
\tau\alpha(Y/\sqrt\tau+A_{\mu+\rho}).
$$
Using the coordinate change $\Phi_\tau$ the integral $I_\mu$ is transformed to
$$
I_\mu(\tau)=\frac{\tau^{\frac{m}{2}}e^{\tau\|\mu+\rho\|^2}}{2^{m-r}}
\int\limits_{\Phi^{-1}(Z_+)}e^{-\|Y\|^2}
\prod\limits_{\alpha\in\Sigma^+}
\alpha(\Phi_\tau(Y)/\tau)^{\frac{m_\alpha}{2}}
(1-e^{-4\alpha(\Phi_\tau(Y))})^{\frac{m_\alpha}{2}}dY
$$ 
Let $\chi_\tau(Y)$ be the characteristic function of the set $\Phi^{-1}(Z_+)$
and let
$$
g_\tau(Y):=\chi_\tau(Y)\prod\limits_{\alpha\in\Sigma^+}
\alpha(Y/\sqrt\tau+A_{\mu+\rho})^{\frac{m_\alpha}{2}}
(1-e^{-4\alpha(\Phi_\tau(Y))})^{\frac{m_\alpha}{2}},
$$
that is now defined on the entire space $Z$ and
$$
I_\mu(\tau)=\frac{\tau^{\frac{m}{2}}e^{\tau\|\mu+\rho\|^2}}{2^{m-r}}
\int\limits_Z e^{-\|Y\|^2}g_\tau(Y)dY.
$$
We want to show that the integral here has a limit as $\tau\ra\infty$. 
First we prove this for the function $g_\tau(Y)$.
\proclaim{Claim} For all $Y\in Z$
$$
\lim\limits_{\tau\ra\infty}g_\tau(Y)=
\cases 
\prod\limits_{\alpha\in\Sigma^+}
\langle\mu+\rho,\alpha\rangle^{\frac{m_\alpha}{2}},& A_{\mu+\rho}\in Z_+\\
0, &A_{\mu+\rho}\in Z\setminus {Z_+}
\endcases
$$
\endproclaim

\demo{Proof of the Claim}
 First let   $A_{\mu+\rho}\in Z_+$. Then $\alpha( A_{\mu+\rho})>0$, 
for all $\Sigma^+$. Let $Y\in Z$ be arbitrary. 
Then  with an appropriate $\tau_0$,
  $\alpha(\sqrt\tau Y+\tau A_{\mu+\rho})>0$ holds
for every $\tau\ge\tau_0$. Thus 
$Y\in\Phi^{-1}_\tau( Z_+)$ and so $\chi_\tau(Y)=1$ for $\tau\ge\tau_0$. Also 
$$
\lim\limits_{\tau\ra\infty}\alpha(Y/\sqrt\tau+A_{\mu+\rho})=\alpha(A_{\mu+\rho})
=\langle\alpha,\mu+\rho\rangle>0
$$
and hence $\lim\limits_{\tau\ra\infty}\alpha(\Phi_\tau Y)=\infty$.
 All these together prove our claim in this case.

Now let $A_{\mu+\rho}\in Z\setminus Z_+$.  
Suppose  there is an $\alpha\in\Sigma^+$ with
 $\alpha(A_{\mu+\rho})<0$. Then for
 all $Y$ in $Z$ there exists some $\tau_0>0$ so that for every $\tau\ge\tau_0$,
 $\alpha(\sqrt\tau Y+\tau A_{\mu+\rho})<0$ and consequently 
$Y\not\in\Phi^{-1}_\tau( Z_+)$ implying $\chi_\tau(Y)=0=g_\tau(Y)$. 
 
Now assume there is at least one $\alpha\in\Sigma^+$  with
 $\alpha(A_{\mu+\rho})=0$ 
and 
$\beta(A_{\mu+\rho})\ge0$ for all $\beta\in\Sigma^+$. Denote by $\Sigma_{+0}$
those  $\beta\in \Sigma^+$, for which $\beta(A_{\mu+\rho})=0$.
 
Let $Y\in Z$. If there exists a $\beta\in\Sigma_{+0}$ with $\beta(Y)\le0$, then
 $\beta(\sqrt\tau Y+\tau A_{\mu+\rho})\le0$ and so  $\chi_\tau(Y)=0=g_\tau(Y)$
 for all
$\tau>0$. 

 Suppose that for all $\beta\in\Sigma_{+0}$, $\beta(Y)>0$. 
Then for all $\tau>0$ and $\beta\in\Sigma_{+0}$, $\beta(\sqrt\tau Y)=
\beta(\Phi_\tau Y)>0$
 and so $0<1-e^{-4\beta(\Phi_\tau(Y))}<1$. Also just as before:
 with an appropriate $\tau_0$,
  $\beta(\sqrt \tau Y+\tau A_{\mu+\rho})>0$ holds
for every $\tau\ge\tau_0$
and $\beta\in\Sigma^+\setminus\Sigma_{+0}$. Thus for all 
$\tau\ge\tau_0$,  $\Phi_\tau(Y)\in Z_+$ hence 
$$
\chi_\tau(Y)=1\quad\text{and}\quad0<\prod\limits_{\alpha\in\Sigma^+}
(1-e^{-4\alpha(\Phi_\tau(Y)})^{\frac{m_\alpha}{2}}<1.
$$
But
$$
\lim\limits_{\tau\ra\infty}\prod\limits_{\alpha\in\Sigma^+}
(\alpha(Y/\sqrt\tau+A_{\mu+\rho}))^{\frac{m_\alpha}{2}}=
\prod\limits_{\alpha\in\Sigma^+}(\alpha(A_{\mu+\rho}))^{\frac{m_\alpha}{2}}=0,
$$
proving that $\lim\limits_{\tau\ra\infty}g_\tau(Y)=0$.
\qed
\enddemo
Now to finish the proof of the theorem we estimate $g_\tau(Y)$.
By its definition
 $g_\tau(Y)$ vanishes outside of the set $\Phi^{-1}(Z_+)$.

 Hence  the trivial estimate yields

$$|g_\tau(Y)|\le\prod\limits_{\alpha\in \Sigma^+}\|\alpha\|^{\frac{m_\alpha}{2}}(\|Y\|+
\|A_{\mu+\rho}\|)^{\frac{m_\alpha}{2}}=:C.
$$

Valid for all $Y\in Z$ and  $\tau\ge1$.   Thus $Ce^{-\|Y\|^2}$ 
is an integrable majorant of $g_\tau(Y)$
for all $\tau\ge1$.   Using Lebesgue's dominated convergence theorem together
with our 
claim and the fact that $\int\limits_Z e^{-\|Y\|^2}dY=\pi^{\frac{r}{2}}$
 finishes the proof of the theorem.
\qed
\enddemo

Back to symmetric spaces again,
let $(M^m=U/K,g)$ be a compact, irreducible, simply connected, Riemannian 
 symmetric space, $\delta$ an irreducible unitary
$K-$sphe\-ri\-cal representation of $U$ with highest weight  $\lambda$.
$\bold c$ denotes Harish-Chandra's $\bold c-$func\-ti\-on
associated to the dual
symmetric space $X=G/K$ and $q_\delta$ is from (2.2.6).
 
\proclaim{Theorem 5.2}
$$
q_\delta(\tau)=2^{r-m}\pi^{\frac{r}{2}}\bold c
(-i\lambda-i\rho)\left(\prod\limits_{\alpha\in\Sigma^+}
\langle\lambda+\rho,\alpha\rangle^{\frac{m_\alpha}{2}}\right)\tau^{\frac{m}{2}}
e^{\tau\|\lambda+\rho\|^2}
(1+o(1)),
$$
as $\tau\ra\infty$.
\endproclaim

\demo{Proof} It follows from
 the Cartan-Helgason theorem ([He2, Theorem 4.1, p.535]), that 
$A_\lambda\in\overline{\fa}_+$. But then Proposition 3.2 implies
with $l=\lambda+\rho$, that
 $A_l\in\fa_+$.
Thus if $\mu$ is a weight of $\delta$, different from $\lambda$,
Proposition 3.1 and Theorem 5.1 (with
$Z=\fa_0$ and $\Sigma^+$ the set 
of positive restricted roots)
yields
$I_\mu(\tau)=I_\lambda(\tau)o(1)$, as $\tau\ra\infty$.
Now using (3.1) for the spherical function corresponding to $\delta$
we get
$$
q_\delta(\tau)=\sum\limits_{\mu\in\Lambda(\delta)}
\langle w_\mu,w_\mu\rangle I_\mu(\tau)\tag5.2
$$  
The discussion above implies, that  
$I_\lambda(\tau)$ dominates all the other terms in (5.2). 
Therefore   (3.2) and Theorem 5.1 finish
the proof.
\qed
\enddemo

Since $\bold c(-i\rho)=1$, Theorem 2.1.1 and Theorem 5.2 yield Theorem 5.3.

\proclaim{Theorem 5.3} If the corrected field of quantum Hilbert spaces
$H^{corr}\ra S$ is projectively flat, then
 for every  irreducible $K-$spherical
representation $\delta$   with  highest weight $\lambda$,
$$
A_\delta=\frac{\bold c(-i\lambda-i\rho)\prod\limits_{\alpha\in\Sigma^+}
\langle\lambda+\rho,\alpha\rangle^{\frac{m_\alpha}{2}}}{
\prod\limits_{\alpha\in\Sigma^+}\langle\rho,\alpha\rangle^{\frac{m_\alpha}{2}}}.\tag5.3
$$
and
$$
B_\delta=\|\lambda+\rho\|^2-\|\rho\|^2.\tag5.4
$$
\endproclaim
Denote by $\Sigma_0$ the set of indivisible restricted roots, i.e. those
$\alpha\in\Sigma$, for which $c\alpha\in\Sigma$ implies $c=\pm1,\pm2.$
 Let $\Sigma^+_0=\Sigma_0\cap\Sigma^+$. As before, for an $\alpha\in\Sigma$
 we take $m_{2\alpha}=0$ if $2\alpha\not\in\Sigma$ and
 $\alpha_0=\alpha/\langle\alpha,\alpha\rangle$.
 %and use the notations of 1d for the root system $\Sigma$.
Now combining Theorem 4.2.1 with Theorem 5.3 we get. 
\proclaim{Theorem 5.4}
Assume the corrected field of quantum Hilbert spaces
$H^{corr}\ra S$ is projectively flat.
Let $\delta$ be an irreducible $K-$spherical
representation
with  highest weight $\lambda$.   Then $A_\delta$ must be equal to $1$,
hence the quantity
$$
\bold c(-i\lambda-i\rho)\prod\limits_{\alpha\in\Sigma_0^+}
\langle\lambda+\rho,\alpha_0\rangle^{\frac{m_\alpha+m_{2\alpha}}{2}}
\tag5.5
$$
is independent of $\delta$ and
$$
\|\lambda+\rho\|^2-\|\rho\|^2=\frac{m}{2}b_\delta,\tag5.6
$$
where $b_\delta$ is from (4.2.2). 
\endproclaim
\demo{Remarks}
1) Proposition 4.2.1 shows
that when $M=U/K$  is  any compact, irreducible, simply connected
Riemannian symmetric space of rank-1, (5-6)
holds for every  irreducible $K-$spherical
representation of $U$. Thus  the  constants $B_\delta$ from Theorem 2.1.1 (b)
do not help in deciding whether the field $H^{corr}\ra S$ is projectively flat
or not.
  It is not clear whether (5-6)  should always hold for the
  higher rank symmetric
  spaces as well, regardless
of projective flatness.

2) If $M$ is isometric to  a compact Lie group $U$ equipped
with a biinvariant metric,
 we know from  [L-Sz3, Theorem 11.3.1] that $H^{corr}\ra S$ is  flat.
Also  it is well known in this case, that for all $\alpha\in\Sigma$, 
$m_\alpha=2$ and $m_{2\alpha}=0$ (i.e. $\Sigma$ is reduced). 
 Now with 
$$
\pi(\nu):=
\prod\limits_{\alpha\in\Sigma^+}\langle\nu,\alpha\rangle,\quad \nu\in\fa_0^*,
$$
we have
$$
\bold c(\nu)=\frac{\pi(\rho)}{\pi(i\nu)}
$$ 
(see [He2, p. 447.]) and the quantity in (5.5) is equal to  $\pi(\rho)$, indeed
independent of $\delta$.
\enddemo

 Next we express  condition  (5.5)
 purely in terms of the root system $\Sigma$ and its multiplicities.
\proclaim{Theorem 5.5} Let $\delta$ be an irreducible $K-$spherical
representation
with  highest weight $\lambda$.
Suppose the corrected field of quantum Hilbert spaces
$H^{corr}\ra S$ is projectively flat. Then the quantity
$$Q(\delta):=
\prod\limits_{\alpha\in\Sigma^+_0}
\frac{\Gamma(\frac1{4}m_\alpha+\frac1{2}
\langle\lambda+\rho,\alpha_0\rangle)
\Gamma(\langle\lambda+\rho,\alpha_0\rangle)                            
\langle\lambda+\rho,\alpha_0\rangle^{\frac{m_{\alpha}+m_{2\alpha}}{2}}}{
\Gamma(\frac1{2}m_\alpha+
\langle\lambda+\rho,\alpha_0\rangle)\Gamma(\frac1{4}m_\alpha+
\frac1{2}m_{2\alpha}+
\frac1{2}\langle\lambda+\rho,\alpha_0\rangle)}
\tag5.7
$$
  is independent of $\delta$.
\endproclaim
If $m_{2\alpha}=0$ and $m_\alpha=2$ for all $\alpha\in\Sigma^+_0$, then it is
obvious that $Q(\delta)$ is in fact independent of $\delta$. This is the group
manifold case.

\demo{Proof}  The 
 Gindikin-Karpelevi\v{c} formula  expresses
Harish-Chandra's $\bold c-$function as a meromorphic function on $\fa^*_\bC$
(see [He2, p.447]), 
$$
\bold c(\nu)=c_0\prod\limits_{\alpha\in\Sigma^+_0}
\frac{2^{\langle-i\nu,\alpha_0\rangle}\Gamma(\langle i\nu,\alpha_0\rangle)}
{\Gamma(\frac1{2}(\frac1{2}m_\alpha+1+
\langle i\nu,\alpha_0\rangle))\Gamma(\frac1{2}(\frac1{2}m_\alpha+m_{2\alpha}+
\langle i\nu,\alpha_0\rangle))}.\tag5.8
$$
Here the constant $c_0$ is determined by $\bold c(-i\rho)=1$.
Using the duplication formula
$$
\Gamma(2z)=2^{2z-1}\pi^{-1/2}\Gamma(z)\Gamma(z+\frac1{2}),
$$
from (5.8) we get
$$
c(-i\lambda-i\rho)=c_1\prod\limits_{\alpha\in\Sigma^+_0}
\frac{\Gamma(\frac1{2}(\frac1{2}m_\alpha+
\langle \lambda+\rho,\alpha_0\rangle))\Gamma(\langle\lambda+
\rho,\alpha_0\rangle)}
{\Gamma(\frac1{2}m_\alpha+
\langle \lambda+\rho,\alpha_0\rangle)
\Gamma(\frac1{2}(\frac1{2}m_\alpha+m_{2\alpha}+
\langle \lambda+\rho,\alpha_0\rangle))},\tag5.9
$$
where
$$
c_1=c_0\prod\limits_{\alpha\in\Sigma^+_0}\frac{2^{m_\alpha/2}}{2\sqrt\pi}.
$$
From (5.5) and (5.9) we see 
(since $(2\alpha)_0=\alpha_0/2$),  that the quantity in (5.5)
does not depend on $\delta$ iff $Q(\delta)$ is independent of $\delta$.   
\qed
\enddemo

\head 6. Root systems and the proof of
Theorem 0.1
\endhead

\subhead 6.1. $\Gamma$-related functions
\endsubhead

\medskip

Here we  take  a closer look at the functions appearing in (5.7)
to find out  which compact symmetric spaces have the property that 
 $Q(\delta)$ (from (5.7))  is   independent of $\delta$.

Let $0<a$, $0\le b,c,d$ be given constants,
$P:=\{z\in\bC : 0<\text{Re z}\}$ and
$$
F(z,a,b,c,d):=\frac{\Gamma(cz+a+b)\Gamma(2cz+2a)(2cz+2a)^{2b+d}}
{\Gamma(2cz+2a+2b)\Gamma(cz+a+b+d)},\tag6.1.1
$$
considered as a function of $z$,
where $\Gamma$ denotes the usual $\Gamma$ function.

\proclaim{Proposition 6.1.1} $F(z,a,b,c,d)$ is a bounded holomorphic function
in a neighborhood of $\overline P$. %and
%$$
%F(z,a,\frac1{2},c,0)\equiv 1.\tag6.1.1
%$$
\endproclaim
\demo{Proof}
Since $\Gamma$ is zero free and holomorphic in $P$, $F$ will be 
holomorphic 
in a neighborhood of $\overline P$.
The substitution $w=cz$ shows that it is enough to prove boundedness when $c=1$.
Let $0<A$ be arbitrary. From
$$
\Gamma(w+A)\sim w^A\Gamma(w),\quad w\ra\infty,\quad w\in P,
$$
(see [Re, p.59]) we get
$$
F(w,a,b,1,d)\sim \frac{w^{a+b}(2w)^{2a}(2w+2a)^{2b+d}}
{{(2w)^{2a+2b}w^{a+b+d}}}\sim 2^d,\quad
w\ra\infty,\quad w\in P,
$$
showing the boundedness of $F$.% and  (6.1.1) follows from the identity
%$\Gamma(w+1)=w\Gamma(w)$.
\qed
\enddemo

Let $0<a_j, c_j$, $0\le b_j, d_j$, $j=1,\dots,N$ and 
$G(z):=\prod\limits_{j=1}^NF(z,a_j,b_j,c_j,d_j)$.

\proclaim{Proposition 6.1.2} Assume that for some $s$, 
$\frac{a_s}{c_s}<\frac{a_j}{c_j}$, for all
 $j\not=s$ and there exists a constant $D\not=0$ with
$G(n)=D$ for all $n\in\bZ_+$. Then $2b_s+d_s=1$.
\endproclaim
\demo{Proof} After renumbering we can assume that $s=1$.
From  Proposition 6.1.1 we know that $G$ is a bounded holomorphic function 
in a neighborhood of $\overline P$. In light of Carlson's theorem 
([T., p.186]), our 
assumptions imply that $G\equiv D$ and so
$$\multline
(2c_1z+2a_1)^{2b_1+d_1}\prod\limits_{j=2}^N(2c_jz+2a_j)^{2b_j+d_j}\\
\equiv 
D\prod\limits_{j=1}^N\frac{\Gamma(2c_jz+2a_j+2b_j)
\Gamma(c_jz+a_j+b_j+d_j)}{\Gamma(c_jz+a_j+b_j)\Gamma(2c_jz+2a_j)}.
\endmultline
$$

Since $a_1/c_1<a_j/c_j$, $1<j$ and
because $\Gamma$ is zero free and holomorphic in
 $\bC\setminus\{0,-1,-2,\dots\}$
 and has first order poles
in the nonpositive integers, 
 the right hand side 
is holomorphic in a neighborhood $U$ of $\{\text{Re} z\ge-\frac{a_1}{c_1}\}$ 
and has a simple zero at 
$-\frac{a_1}{c_1}$. Furthermore
  $\prod\limits_{j=2}^N(2c_jz+2a_j)^{2b_j+d_j}$ is   holomorphic and zero free
in $U$. Hence $(2c_1z+2a_1)^{2b_1+d_1}$ should extend 
holomorphically to a neighborhood of $z_0:=-\frac{a_1}{c_1},$ with a first
order zero at $z_0$.
But this  happens iff $2b_1+d_1=1$.
\qed
\enddemo

\subhead 6.2 Root systems
\endsubhead

Let $(Z,\langle.,.\rangle)$ be an $r-$dimensional Euclidean space. For
$0\not=\alpha\in Z$   let
 $\alpha_0=\alpha/\langle\alpha,\alpha\rangle$. %Then $(2\alpha)^\vee=\alpha_0$.

Let $R\subset Z$ be a (possible nonreduced) 
 root system.
Choose a basis  
$\alpha_1,\dots,\alpha_r$  of $R$ and let $R^+$ be the set of
positive roots, $\Bbb Z_+:=\{0,1,2,\ldots\}$.
$$
P_+:=\{\gamma\in Z : \langle\gamma,\alpha_0\rangle
\in \bZ_+ , \forall \alpha\in R^+\}.\tag6.2.1
$$
According to the Cartan-Helgason theorem ([He2, Theorem 4.1, p.535,
Corollary 4.2, p.538]),
when $Z=\fa_0^*$ and $R=\Sigma$ the set of restricted roots of a compact, simply
connected Riemannian 
symmetric space $M=U/K$,
the highest weights
of the irreducible $K$-spherical representations of $U$ are precisely
the elements of $P_+$.

A  multiplicity function on $R$ is a map $m:R\ra \bR$,
denoted by $\alpha\mapsto m_\alpha$  such that $m_{w\alpha}=m_\alpha$
 for every Weyl group element $w$.
Let  
$\rho:=\frac1{2}\sum\limits_{\alpha\in R_+}m_\alpha\alpha$.
Denote by $R_0$ the set of indivisible roots and
$R_0^+=R^+\cap R_0$.
Inspired by the formula (5.7) for $Q(\delta)$, we define the analogous
function for  $\mu\in P_+$ as follows.
$$Q(\mu):=
\prod\limits_{\alpha\in R^+_0}
\frac{\Gamma(\frac1{4}m_\alpha+\frac1{2}
\langle\mu+\rho,\alpha_0\rangle)
\Gamma(\langle\mu+\rho,\alpha_0\rangle)                            
\langle\mu+\rho,\alpha_0\rangle^{\frac{m_{\alpha}+m_{2\alpha}}{2}}}{
\Gamma(\frac1{2}m_\alpha+
\langle\mu+\rho,\alpha_0\rangle)\Gamma(\frac1{4}m_\alpha+
\frac1{2}m_{2\alpha}+
\frac1{2}\langle\mu+\rho,\alpha_0\rangle)}
\tag6.2.2
$$
(6.2.5) shows that this is a well defined quantity when all multiplicities
are positive.
Denote by $R_*$ the set of unmultipliable roots in $R$.
A basis $\beta_1,\dots,\beta_r$ of $R_*$ can be obtained
by taking $\beta_j=\alpha_j$ if $2\alpha_j\not\in R$ and $\beta_j=2\alpha_j$
if $2\alpha_j\in R$.
Define
$\mu_j\in Z$, $j=1,\dots,r$  by
$$
\langle\mu_j,\beta_{k,0}\rangle=
\delta_{jk},\quad j,k=1,\dots,r.\tag6.2.3
$$
Then
$$
\mu\in P_+\quad \text{if and only if}\quad \mu=\sum\limits_{j=1}^rn_j\mu_j\quad
\text{with}\quad n_j\in  \bZ_+\tag6.2.4
$$
([He3, Proposition 4.23, p.150]).

\proclaim{Proposition 6.2.1}
Suppose that $0<m_\alpha$ for all $\alpha\in R$. Then
$$
0<\langle\rho,\alpha\rangle\quad\text{and}\quad
0\le\langle\mu,\alpha\rangle\quad \forall\alpha\in R^+
,\forall\mu\in P_+.
\tag6.2.5
$$
For a fixed  $1\le j\le r$, let
$R^+_j:=\{\alpha\in R^+_0 : 0<\langle\mu_j,\alpha_0\rangle\}$.
Then  
$$
\frac{\langle\rho,\alpha_{j,0}\rangle}{\langle\mu_j,\alpha_{j,0}\rangle}
< \frac{\langle\rho,\alpha_0\rangle}{\langle\mu_j,\alpha_0\rangle},
\quad\forall\alpha\in R^+_j,\quad \alpha\not=\alpha_j.
\tag6.2.6
$$
\endproclaim
\demo{Proof} The proof of Proposition 3.2 also works here, showing the first
part of (6.2.5). The second  part follows from (6.2.3) and (6.2.4). 
If $\alpha_j\not=\alpha\in R^+_j$, then $\alpha=\sum\limits^r_1n_s\alpha_s$ with 
$n_s\in\bZ_+$. From (6.2.3) we have $0<\langle\mu_j,\alpha_j\rangle$ and
$$
0<\langle\mu_j,\alpha\rangle=n_j\langle\mu_j,\alpha_j\rangle.\tag6.2.7
$$
Hence $0<n_j$. Since $\alpha$ is indivisible and is different from
$\alpha_j$, there must be at least one
more $s$ with $0<n_s$. (6.2.5) then implies
$$
\langle\rho,n_j\alpha_j\rangle<\langle\rho,\alpha\rangle.\tag6.2.8
$$
Now in light of (6.2.7), if we divide (6.2.8) by
$n_j\langle\mu_j,\alpha_j\rangle$
we get (6.2.6).
\qed
\enddemo

 We call a multiplicity function  $m:R\ra\bR$
{\it geometric} if it takes only positive integer
values and satisfies the following property: if  $\alpha\in R$ and
$m_\alpha$ is odd, then $2\alpha\not\in R$. For $\alpha\in R$
we use the convention as before: $m_{2\alpha}=0$ if $2\alpha$ is not a root.
If $R=\Sigma$, a restricted root system of a compact, Riemannian
symmetric space,
its multiplicity function is geometric in this sense, see [Ar, Proposition 2.3]
or [He1, p.530, 4F].

\proclaim{Theorem 6.2.2} Let $R$ be an irreducible  root system
with  a geometric multiplicity
function $m$. 
Suppose  $Q(\mu)$, $\mu\in P_+$  
is independent of $\mu$ ($Q(\mu)$ is from (6.2.2)).
Then $R$ is reduced and 
for all $\alpha\in R$, $m_\alpha=2$.
\endproclaim
\demo{Proof}
Let $\beta_j\in R$, $\mu_j\in Z$ as in (6.2.3) and fix a $j$ with $1\le j\le r$.
From (6.2.3) we have $n\mu_j\in P_+$ for all $n\in\bZ_+$.
Now let
$$
H_j(z):=\prod\limits_{\alpha\in R^+_0}
F\left(z,\frac{\langle\rho,\alpha_0\rangle}{2},
\frac{m_\alpha}{4},\frac{\langle\mu_j,\alpha_0\rangle}{2},
\frac{m_{2\alpha}}{2}\right),
$$
where $F$ is from (6.1.1). Then from (6.2.2) we get
$$
Q(n\mu_j)=H_j(n),\quad\forall n\in \bZ_+.
$$
By our assumption on $Q$,  $H_j(n)$ will be independent of $n$.
For any values of the parameters $a, b, d$,
the function $F(z,a,b,0,d)$ from (6.1.1) is always a nonzero constant.
Thus if we leave out from the definition of $H_j$ all those terms
that correspond to a root  $\alpha\in R^+_0$
with $\langle\mu_j,\alpha_0\rangle=0$, the result is still a function
that is a nonzero constant on the nonnegative integers. Let
$R^+_j:=\{\alpha\in R_0^+ : \langle\mu_j,\alpha_0\rangle>0\}$
be as in Proposition 6.2.1 and 
$$
G_j(z):=\prod\limits_{\alpha\in R^+_j}
F\left(z,\frac{\langle\rho,\alpha_0\rangle}{2},
\frac{m_\alpha}{4},\frac{\langle\mu_j,\alpha_0\rangle}{2},
\frac{m_{2\alpha}}{2}\right).
$$
Then we still have that 
$G_j(n)$ is a nonzero constant when $n\in\bZ_+$. This together with
(6.2.6) and Proposition 6.1.2 implies 
$$
m_{\alpha_j}+m_{2\alpha_j}=2.\tag6.2.9
$$
Since $m$ is a geometric multiplicity function,  (6.2.9) yields 
$m_{\alpha_j}=2$ and $m_{2\alpha_j}=0$. Thus $2\alpha_j$ is not a root. Since $R_0$,
$R$ and
$m$ are
Weyl group invariant, this yields that
$R$ is reduced and $m\equiv2$.
\qed
\enddemo
\subhead  Proof of Theorem 0.1
\endsubhead
If $(M,g)$ is an irreducible, simply connected, compact,
 Riemannian symmetric  space,
%$H^{corr}\ra S$ is projectively flat.
 the  set of restricted roots $\Sigma$ in $\fa_0^*$
forms an irreducible root system with a geometric multiplicity function.
In light of Theorem 5.5 and Theorem 6.2.2,
projective flatness of $H^{corr}\ra S$ 
implies 
$\Sigma$ is reduced and all the multiplicities are equal to $2$. 
 But these conditions characterize compact Lie groups among
compact Riemannian symmetric spaces ([L, Theorem 4.4, p.82]).
\qed

{\bf Acknowledgments.} Most of this research was done while I was visiting
the Alfr\'ed R\'enyi Institute of Mathematics. I would like to thank the
institute for its hospitality and  financial support.
 This research was also partially supported by OTKA grant K112703.
 I thank the anonymous referee for his/her very careful reading
 of the paper and the numerous comments and
 suggestions which significantly contributed to improving
 the presentation of this work.

\enddocument

\Refs
\widestnumber\key{XXXX}
\ref\key A
\by R. M. Aguilar
\paper Symplectic reduction and the homogeneous complex Monge-Amp\'ere equation
\jour Ann. Glob. Anal. Geom.
\vol 19
\yr 2001
\pages 327-353
\endref


\ref\key Ar
\by S. Araki
\paper On root systems and an infinitesimal classification of irreducible
symmetric spaces
\jour J. Math. Osaka City Univ.
\vol 13
\pages 1-34
\yr 1962
\endref


\ref\key ADW\by S.~Axelrod, S.~Della Pietra, E.~Witten\paper Geometric 
quantization of Chern--Simons gauge theory
\jour J.~Diff.~Geo.\vol33\yr 1991\pages 787--902
\endref


\ref\key Bl1\manyby\ R.J.~Blattner
\paper Quantization and representation theory
\inbook Proc.~Symp.~Pure Math.\vol 26
\pages 147--165\publ Amer.~Math.~Soc.\publaddr Providence\yr 1973\endref



\ref\key Bl2\bysame\paper The meta--linear geometry of non--real polarizations
\inbook Lecture Notes in Math.\vol 570 (1975)
\pages 11--45\publ Springer\publaddr Berlin\yr 1977
\endref


\ref\key Ch
\by L. Charles
\paper Semi-classical properties of geometric quantization with metaplectic 
correction
\jour Commun. Math. Phys.
\vol 270
\pages 445-480
\yr 2007
\endref

\ref
\key F
\by J. Faraut
\paper Espaces Hilbertiens invariant de fonctions holomophes
\inbook Semin. Congr. Vol.7. Soc. Math. de France, Paris
\yr 2003
\pages 101-167
\endref


\ref\key FMN1\manyby C.~Florentino, P.~Matias, J.~Mour\~ao, 
J.~Nunes\paper Geometric quantization, complex structures and the coherent 
state transform\jour J.~Funct.~Anal.\vol 221\yr 2005\pages 303--322\endref





\ref\key FMN2\bysame\paper On the BKS pairing for K\"ahler quantizations of 
the cotangent bundle of a Lie group\jour J.~Funct.~Anal.\vol234\yr 2006
\pages 180--198\endref



\ref\key FU\manyby T.Foth, A. Uribe
\paper The manifold of compatible almost complex structures and geometric
quantization
\jour Commun. Math. Phys.
\vol 274
\yr 2007
\pages 357-379
\endref



\ref\key GS\by V.~Guillemin, M.~Stenzel\paper Grauert tubes and the 
homogeneous Monge--Amp\`ere equation\jour J.~Diff.~Geom.\vol34
\yr 1991\pages 561--570\endref



\ref\key Hal1\manyby B.C.~Hall\paper 
The Segal--Bargmann ``coherent state'' transform for compact Lie groups
\jour J.~Funct.~Anal.\vol 122\yr 1994\pages 103--151\endref





\ref\key Hal2\bysame\paper Geometric quantization and the generalized 
Segal--Bargmann transform for Lie groups of compact type
\jour Comm.~Math.~Phys.\vol226\yr 2002\pages 233--268\endref




\ref\key HK
\by B. C. Hall, W. D. Kirwin
\paper Adapted complex structures and the geodesic flow
\jour Math. Ann.
\vol 350
 \issue 2
\pages 455-474 
\yr 2011
\endref


\ref
\key Ha1
\by Harish-Chandra
\paper Spherical functions on a semisimple Lie group I
\jour Amer. J. Math
\vol 80
\issue 2
\yr 1958
\pages 241-310
\endref





\ref
\key Ha2
\bysame
\paper Spherical functions on a semisimple Lie group I
\jour Amer. J. Math
\vol 80,
\issue 3
\yr 1958
\pages 553-613
\endref



\ref
\key H
\by G. J. Heckmann
\paper Root systems and hypergeometric functions  II
\jour Comp. Math
\vol 64
\yr 1987
\pages 353-373
\endref








\ref
\key HO1
\by G. J. Heckmann, E. M. Opdam 
\paper Root systems and hypergeometric functions  I
\jour Comp. Math
\vol 64
\yr 1987
\pages 329-3352
\endref



\ref
\key HO2
\bysame
\paper Jacobi polynomials and hypergeometric functions associated with
root systems
\paperinfo http://www.math.ru.nl/$\sim$heckman/
\yr 2015
\endref




\ref\key HS
\by G. J. Heckmann, H. Schlicktkrull
\book Harmonic analysis and special functions on symmetric spaces
\bookinfo Perspectives in Mathematics, Vol.16
\publ Academic Press
\publaddr San Diego
\yr 1994
\endref

\ref\key He1\manyby S.~Helgason\book Differential geometry,
Lie groups, and symmetric spaces\publ Amer. Math. Soc.
\publaddr Providence\yr 2001\endref


\ref\key He2\bysame\book Groups and geometric analysis, integral
geometry, invariant differential operators and spherical functions
\publ Amer. Math. Soc.
\publaddr Providence\yr 2002\endref 

\ref\key He3\bysame\book Geometric Analysis on Symmetric Spaces
\publ Amer. Math. Soc.
\publaddr Providence\yr 1994\endref


\ref\key Hi\by N.~Hitchin\paper Flat connections and geometric quantization
\jour Comm.~Math.~Phys.\vol 131\yr 1990\pages 347--380\endref



%\ref
%\key Hu
%\by J. E. Humphreys
%\book Introduction to Lie algebras and representation theory
%\publ Springer-Verlag
%\publaddr New York
%\yr 1972
%\endref

\ref
\key Hu
\by J. E. Humphreys
\book Reflection groups and Coxeter groups
\publ CUP
\yr 1990
\endref


\ref\key Ko1\manyby B.~Kostant\paper Quantization and unitary representations
 I.\inbook
Lectures in modern analysis and applications III, 
Lecture Notes in Math.\vol 170\publ Springer\publaddr Berlin\yr 1970
\pages 87--208\endref




\ref\key Ko2\bysame\paper Symplectic spinors\inbook Symposia Mathematica XIV
\pages 139--152\publ Academic Press\publaddr London\yr 1974
\endref




\ref\key KW\by W.D.~Kirwin, S.~Wu\paper Geometric quantization, parallel 
transport and the Fourier transform\jour Comm.~Math.~Phys.\vol266\yr 2006
\pages 577--594\endref



\ref\key LSz1
\by L. Lempert, Sz\H oke
\paper Global solutions of the homogeneous complex Monge-Amp\'ere equation
and complex structures on the tangent bundle of Riemannian manifolds
\jour Math. Ann
\vol 290
\yr 1991
\pages 689-712
\endref





\ref\key LSz2
\bysame
\paper A new look at adapted complex structures
\jour Bull. Lond. Math. Soc
\vol 44
\yr 2012
\pages 367-374
\endref



\ref\key LSz3
\bysame 
\paper Direct images, fields of Hilbert spaces, and geometric quantization
\jour Commun. Math. Phys.
\vol 327
\yr 2014
\pages 49-99
\endref

\ref
\key LSz4
\bysame
\paper Curvature of fields of quantum Hilbert spaces
\jour Quart. J. Math.
\vol 66
\paperinfo doi: 10.1093/qmath/hau037
\yr 2015
\pages 645-657
\endref

\ref
\key L
\by O. Loos
\book Symmetric spaces II: compact spaces and classification
\publ  Benjamin
\publaddr New York, Amsterdam
\yr 1969
\endref


\ref
\key M
\by J. D. Murray
\book Asymptotic analysis
\publ Springer-Verlag
\publaddr New York
\yr 1984
\endref



\ref
\key OW
\by G. \'Olafsson, K. Wiboonton
\paper The Segal-Bargmann transform on compact symmetric spaces and their 
direct limits
\inbook 
Progress in Mathematics, Vol 306: Lie groups: Structure, actions, and
representations
\bookinfo In Honor of J. A. Wolf on the occasion of his 75th birthday
\eds A. Huckleberry, I. Penkov, G. Zuckerman
\publ Birkh\"auser
\yr 2013
\endref



\ref\key R
\by H. Rawnsley
\paper A non-unitary pairing of polarizations for the Kepler problem 
\jour Trans. Amer. Math. Soc.
\vol 250
\yr 1979
\pages 167-180
\endref

\ref\key Re
\by R. Remmert
\book Classical topics in complex function theory
\publ Springer-Verlag
\publaddr New York
\yr 1998
\endref

\ref\key So\by J.-M.~Souriau\book Structure des syst\`emes dynamiques
\publ Dunod\publaddr Paris\yr 1970\endref






\ref
\key S
\by M. Stenzel
\paper The Segal-Bargmann transform on a symmetric space of compact type
\jour J. Funct. Anal.
\vol 195
\yr 1999
\pages 44-58
\endref

\ref\key Sz1\by R.~Sz\H{o}ke\paper Complex structures on 
tangent bundles of Riemannian manifolds\jour Math.~Ann.
\vol 291\yr 1991\pages 409--428\endref

\ref\key Sz2\bysame
\paper Automorphisms of certain Stein manifolds
\jour Math. Z.
\vol 219
\pages 357-385
\yr 1995
\endref

\ref\key Sz3\bysame\paper Adapted complex structures and 
Riemannian homogeneous spaces\jour Ann.~Polon.
Math.~LXX\yr 1998\pages 215--220\endref

\ref\key Sz4
\bysame
\paper Smooth structures on the field of prequantum Hilbert spaces
\paperinfo preprint
\endref



\ref
\key T
\by E. C. Titchmarsh
\book The theory of functions,{\ \rm 2nd ed.}
\publ OUP
\publaddr London
\yr 1939
\endref


\ref\key Vi
\by A. Vi\~na
\paper Identification of K\"ahler quantizations and the Berry phase
\jour J. Geom. Phys.
\vol 36
\yr 2000
\pages 223-250
\endref




\ref
\key V
\by L. Vretare
\paper Elementary spherical functions on symmetric spaces
\jour  Math. Scand
\vol 39
\yr 1976
\pages 343-358
\endref


\ref\key Wo\by N.M.J.~Woodhouse\book Geometric quantization, {\rm 2nd ed.}
\publ Clarendon Press\publaddr Oxford\yr 1992\endref


\endRefs
\bye